\documentclass[12pt,a4paper]{article}
\usepackage[margin=1in]{geometry}
\usepackage{authblk}
\usepackage{fullpage}
\usepackage{amsthm}
\usepackage{graphicx}
\usepackage{url}
\usepackage{xcolor}
\usepackage{natbib}
\usepackage{amsmath}
\usepackage{amssymb}
\usepackage{arydshln}
\usepackage{hyperref}
\usepackage{listings}
\usepackage[ruled]{algorithm2e}
\usepackage{microtype}

\usepackage{subfig}

\graphicspath{{./images/}}

\definecolor{dkgreen}{rgb}{0,0.6,0}
\definecolor{gray}{rgb}{0.5,0.5,0.5}
\definecolor{mauve}{rgb}{0.58,0,0.82}

\newcommand{\mb}{\mathbf}

\newcommand{\Expects}[2]{\mathbb{E}_{{#1}} \left[{#2}\right]}
\newcommand{\Var}[1]{\mbox{Var} \left[{#1}\right]}

\newcommand{\Vars}[2]{\mbox{Var}_{{#1}} \left[{#2}\right]}

\newcommand{\bx}{\mathbf{x}}
\newcommand{\bX}{\mathbf X}
\newcommand{\bD}{\mathbf D}
\newcommand{\bP}{\mathbf P}
\newcommand{\bQ}{\mathbf Q}
\newcommand{\bZ}{\mathbf Z}
\newcommand{\bI}{\mathbf I}
\newcommand{\by}{\mathbf y}
\newcommand{\btheta}{\boldsymbol\theta}
\newcommand{\bV}{\mathbf V}
\newcommand{\bzeta}{\boldsymbol\zeta}

\newcommand{\bnu}{\boldsymbol\nu}
\newcommand{\brho}{\boldsymbol\rho}
\newcommand{\bSigma}{\boldsymbol\Sigma}

\newcommand{\st}{\btheta} %%notation for sample

\newcommand{\CN}[1]{{\color{red} CN: #1}} % Chris's notes 

\begin{document}
\title{Stochastic gradient Markov chain Monte Carlo}

\author{Christopher Nemeth\footnote{c.nemeth@lancaster.ac.uk} }
\author{Paul Fearnhead}

\affil{Department of Mathematics and Statistics, Lancaster University, Lancaster, UK}
\maketitle

%% an abstract and keywords
\begin{abstract}
  Markov chain Monte Carlo (MCMC) algorithms are generally regarded as the gold standard technique for Bayesian inference. They are theoretically well-understood and conceptually simple to apply in practice. The drawback of MCMC is that in general performing exact inference requires all of the data to be processed at each iteration of the algorithm. For large data sets, the computational cost of MCMC can be prohibitive, which has led to recent developments in scalable Monte Carlo algorithms that have a significantly lower computational cost than standard MCMC. In this paper, we focus on a particular class of scalable Monte Carlo algorithms, stochastic gradient Markov chain Monte Carlo (SGMCMC) which utilises data subsampling techniques to reduce the per-iteration cost of MCMC. We provide an introduction to some popular SGMCMC algorithms and review the supporting theoretical results, as well as comparing the efficiency of SGMCMC algorithms against MCMC on benchmark examples. The supporting R code is available online\footnote{https://github.com/chris-nemeth/sgmcmc-review-paper}.
\end{abstract}


\section{Introduction}
\label{sec:introduction}

The Bayesian approach to modelling data provides a flexible mathematical framework for incorporating uncertainty of unknown quantities within complex statistical models. %Starting from \textit{a priori} knowledge of the uncertainty in the model parameters, a prior distribution is formed and from this, a posterior distribution, conditional on the existing data, can be derived which accounts for the \textit{a posteriori} parameter uncertainty. 
The Bayesian posterior distribution encodes the probabilistic uncertainty in the model parameters and can be used, for example, to make predictions for new unobserved data. In general, the posterior distribution cannot be calculated analytically and it is therefore necessary to approximate it. Deterministic approximations, such as the Laplace approximation \citep[][see Section 4.4]{bishop2006pattern}, variational Bayes \citep{blei2017variational} and expectation-propagation \citep{minka2001expectation}, aim to approximate the posterior with a simpler tractable distribution (e.g. a normal distribution). These deterministic approximations are often fit using fast optimisation techniques and trade-off exact posterior inference for computational efficiency.

Markov chain Monte Carlo (MCMC) algorithms \citep{brooks2011handbook} are a class of stochastic simulation-based techniques which approximate the posterior distribution with a discrete set of samples. The posterior samples are generated from a Markov chain whose invariant distribution is the posterior distribution. Simple MCMC algorithms, such as random-walk Metropolis \citep{metropolis1953equation}, are easy to apply and only require that the unnormalised density of the posterior can be evaluated point-wise. More efficient MCMC algorithms, which offer faster exploration of the posterior, utilise gradients of the posterior density within the proposal mechanism \citep{Roberts:1996,Neal:2011,girolami2011riemann}.  %For certain model types, where the posterior distribution can be factorised into conditionally conjugate components, samples from the posterior distribution can be generated using Gibbs sampling \citep{GEMA1984} by sequentially sampling from the lower-dimensional conditional distributions. 
Under mild conditions, it is easy to show that asymptotically the samples generated from the Markov chain converge to the posterior distribution \citep{roberts2004general} and for many of the popular MCMC algorithms, rates of convergence based on geometric ergodicity have been established \citep[see][for details]{meyn1994computable,roberts1997geometric}. 

Under weak conditions, MCMC algorithms have the advantage of providing asymptotically exact posterior samples, but at the expense of being computationally slow to apply in practice. This issue is further exacerbated by the demand to store and analyse large-scale data sets and to fit increasingly sophisticated and complex models to these high-dimensional data. For example, scientific fields, such as population genetics \citep{raj2014faststructure}, brain imaging \citep{andersen2018bayesian} and natural language processing \citep{yogatama2014dynamic}, commonly use a Bayesian approach to data analysis, but the continual growth in the size of the data sets in these fields prevents the use of traditional MCMC methods. Computational challenges such as these have led to recent research interest in scalable Monte Carlo algorithms. Broadly speaking, these new Monte Carlo techniques achieve computational efficiency by either parallelising the MCMC scheme, or by subsampling the data.

If the data can be split across multiple computer cores then the computational challenge of inference can be parallelised, where an MCMC algorithm is applied on each core to draw samples from a partial posterior that is conditional on only a subset of the full data. The challenge is then to merge these posterior samples from each computer to generate an approximation to the full posterior distribution. It is possible to construct methods to merge samples that are exact if the partial posteriors are Gaussian \citep{scott2016bayes}; for example with update rules that just depend on the mean and variance for each partial posterior.  However, it is hard to quantify the level of approximation such rules introduce due to non-Gaussianity of the partial posteriors. Alternative merging procedures, that aim to be more robust to non-Gaussianity, have also been proposed \citep{neiswanger2013asymptotically,rabinovich2015variational,srivastava2018scalable,nemeth2018merging}, but it is hard to quantify the level of approximation accuracy such merging procedures have in general.

Alternatively, rather than using multiple computer cores, a single MCMC algorithm can be used, where only a subsample of the data is evaluated at each iteration \citep{bardenet2017markov}. For example, in the Metropolis-Hastings algorithm, the accept-reject step can be approximated with a subset of the full data \citep{korattikara2014austerity,bardenet2014towards,quiroz2018speeding}.  %As in the case of parallel MCMC, subsampling methods introduce an approximation error into the posterior, but this error is often small relative to the computational benefits. 
Again these methods introduce a trade-off between computational speed-up and accuracy. For some models, it is possible to use subsamples of the data at each iteration with the guarantee of sampling from the true posterior; e.g., continuous-time MCMC methods \citep{Fearnhead:2018,Bierkens:2019,bouchard2018bouncy}. This is possible if the derivative of log-posterior density can be globally bounded. 

% SGMCMC - langevin - sgd
Perhaps the most general and popular class of scalable, subsampling-based algorithms are stochastic gradient MCMC methods. These algorithms are derived from the discrete-time approximations of continuous-time diffusion processes. The simplest of these being the over-damped Langevin diffusion \citep{Roberts:1996}, which admits the posterior as its invariant distribution. However, in practice, a discrete-time Euler approximation of the diffusion is used for Monte Carlo sampling, which is known as the unadjusted Langevin algorithm. Due to the discretisation error, samples generated from the unadjusted Langevin algorithm only approximately maintain the posterior as its invariant distribution, which can be made exact using a Metropolis-type correction \citep{besag1994comments}. Even without the Metropolis correction, the unadjusted Langevin algorithm can be computationally expensive as the gradient of the log-posterior density requires the evaluation of the full data. Inspired by stochastic gradient descent \citep{robbins1951stochastic}, \cite{Welling:2011} proposed the stochastic gradient Langevin algorithm, where the gradient component of the unadjusted Langevin algorithm is replaced by a stochastic approximation calculated on a subsample of the full data. An advantage of stochastic gradient MCMC over other subsampling-based MCMC techniques, such as piece-wise deterministic MCMC \citep{Fearnhead:2018}, is that it can be applied to a broad class of models and in the simplest case, only requires that the first-order gradient of the log-posterior density can be evaluated point-wise. A drawback of these algorithms is that, while producing consistent estimates and satisfying a central limit theorem \citep{Teh:2016}, they converge at a slower rate than traditional MCMC algorithms. In recent years, stochastic gradient MCMC algorithms have become a popular tool for scalable Bayesian inference, particularly in the machine learning community, and there have been numerous methodological \citep{ma2015complete,chen2014stochastic,Dubey:2016,Baker:2017} and theoretical developments \citep{Teh:2016,Vollmer:2016,Dalalyan:2017,Durmus:2017} along with new application areas for these algorithms \citep{balan2015bayesian,gan2015,wang2015}. This paper presents a review of some of the key developments in stochastic gradient MCMC and highlights some of the opportunities for future research.

% This paper is organised as follows. 
This paper is organised as follows. Section \ref{sec:langevin-based-mcmc} introduces the Langevin diffusion and its discrete-time approximation as the basis for stochastic gradient MCMC. This section also presents theoretical error bounds on the posterior approximation and an illustrative example of stochastic gradient Langevin dynamics on a tractable Gaussian example. In Section \ref{sec:altern-stoch-grad}, we extend the stochastic gradient MCMC framework beyond the Langevin diffusion to a general class of stochastic differential equations with many popular stochastic gradient MCMC algorithms given as special cases.  Like many MCMC algorithms, stochastic gradient MCMC has tuning parameters which affect the efficiency of the algorithm. Standard diagnostics for tuning traditional MCMC algorithms are not appropriate for stochastic gradient MCMC and Section \ref{sec:diagnostic-tests} introduces the kernel Stein discrepancy as a metric for both tuning and assessing convergence of stochastic gradient MCMC algorithms. Section \ref{sec:extend-stoch-grad} reviews some of the recent work on extending these algorithms to new settings beyond the case where data are independent and the model parameters are continuous on the real space. A simulation study is given in Section \ref{sec:simulation-study}, where several stochastic gradient MCMC algorithms are compared against traditional MCMC methods to illustrate the trade-off between speed and accuracy. Finally, Section \ref{sec:discussion} concludes with a discussion of the main points in the paper and highlights some areas for future research.

\section{Langevin-based Stochastic Gradient MCMC}
\label{sec:langevin-based-mcmc}

\subsection{The Langevin Diffusion}
\label{sec:langevin-diffusion}

We are interested in sampling from a target density $\pi(\st)$, where we assume $\st \in \mathbb{R}^d$ and the unnormalised density is of the form, 
\begin{equation}
  \label{eq:target}
  \pi(\st) \propto \exp\{ -U(\st)\},  
\end{equation}
and defined in terms of a \textit{potential function} $U(\st)$. We will assume that $U(\st)$ is continuous and differentiable almost everywhere, which are necessary requirements for the methods we discuss in this paper. In our motivating applications from Bayesian analysis for big data, the potential will be defined as a sum over data points. For example, if we have independent data, $y_1,\ldots,y_N$ then $\pi(\st)\propto p(\st)\prod_{i=1}^N f(y_i|\st)$, where $p(\st)$ is the prior density and $f(y_i|\st)$ is the likelihood for the $i$th observation. In this setting, we can define $U(\st)=\sum_{i=1}^N U_i(\st)$, where $U_i(\st)= - \log f(y_i|\st)-(1/N)\log p(\st)$.  

One way to generate samples from $\pi(\st)$ is to simulate a stochastic process that has $\pi$ as its stationary distribution. If we sample from such a process for a long time period and throw away the samples we generate during an initial burn-in period, then the remaining samples will be approximately distributed as $\pi$. The quality of the approximation will depend on how fast the stochastic process converges to its stationary distribution from the initial point, relative to the length of the burn-in period. The most common example of such an approach to sampling is MCMC \citep{Hastings:1970}.

Under mild regularity conditions \citep{Roberts:1996,pillai2012optimal}, the Langevin diffusion, defined by the stochastic differential equation
\begin{equation} \label{eq:LangevinSDE}
\mbox{d}\st(t)=-\frac{1}{2} \nabla U(\st(t)) \mbox{d}t+\mbox{d}B_t,
\end{equation}
where $\nabla U(\st(t))$ is the drift term and $B_t$ denotes $d$-dimensional Brownian motion, has $\pi$ as its stationary distribution. This equation can be interpreted as defining the dynamics of a continuous-time Markov process over infinitesimally-small time intervals. That is, for a small time-interval $h>0$, the Langevin diffusion has approximate dynamics given by
\begin{equation} \label{eq:Euler}
    \st(t+h)\approx \st(t) - \frac{h}{2} \nabla U(\st(t)) + \sqrt{h} \bZ, \quad k=0,\ldots,K
\end{equation}
where $\bZ$ is a vector of $d$ independent standard Gaussian random variables. 

%DO WE NEED TO SAY SOMETHING ABOUT THE INITIAL STATE? \CN{We could leave this to the theory section as the impact of the initial state is implicit in the W2 bound}

The dynamics implied by (\ref{eq:Euler}) give a simple recipe to approximately sample from the Langevin diffusion. To do so over a time period of length $T=Kh$, for some integer $K$, we just set $\st_0$ to be the initial state of the process and repeatedly simulate from (\ref{eq:Euler}) to obtain values of the process at times $h,2h,\ldots,Kh$. In the following, when using such a scheme we will use the notation $\st_k$ to denote $\st(kh)$, the state at time $kh$. %(If we wish the value of the process at some time, $t$, which is not a multiple of $h$ we can interpolate between the values simulated at integer multiples of $h$ immediately prior to and after $t$.) Furthermore, 
If we are interested in sampling from the Langevin diffusion at some fixed time $T$, then the Euler discretisation will become more accurate as we decrease $h$; and we can achieve any required degree of accuracy if we choose $h$ small enough. However, it is often difficult in practice to know when $h$ is small enough. 

%If we let $p_t$ to be the density for $\st_t$ simulated from our given initial state under the true Langevin dynamics, and $p_t^{\epsilon}$ to be the density for $\st_t$ simulated under the Euler approximation, then it is natural to ask, how close are these two densities? Under weak regularity conditions, and for a suitable definition of distance, the distance between the distributions characterised by these two densities decays as $h\rightarrow 0$, with the rate of decay being ??.
%see REFS for more details. This means that for any required degree of accuracy, we can choose $h$ small enough that our Euler approximation gives a close enough approximation to the true Langevin diffusion. 
 
\subsection{Approximate MCMC using the Langevin Diffusion}
\label{sec:appr-mcmc-using}

%SOMEWHERE WE MAY NEED TO CLARIFY NOTATION $\st_t$ and $\st_k$ WITH ONE CTS TIME AND THE OTHER DISCRETE. \CN{I've attempted to clarify this with $\st(t)$ and $\st_k$ respectively}

As the Langevin diffusion has $\pi$ as its stationary distribution, it is natural to consider this stochastic process as a basis for an MCMC algorithm. In fact, if it were possible to simulate exactly the dynamics of the Langevin diffusion, then we could use the resulting realisations at a set of discrete time-points as our MCMC output. However, for general $\pi(\st)$ the Langevin dynamics are intractable, and in practice people often resort to using samples generated by the Euler approximation (\ref{eq:Euler}).

This is most commonly seen with the Metropolis-adjusted Langevin Algorithm, or MALA \cite[]{Roberts:1996}. This algorithm uses the Euler approximation (\ref{eq:Euler}) over an appropriately chosen time-interval, $h$, to define the proposal distribution of a standard Metropolis-Hastings algorithm. The simulated value is then either accepted or rejected based on the Metropolis-Hastings acceptance probability. Such an algorithm has good theoretical properties, and in particular, can scale better to high-dimensional problems than the simpler random walk MCMC algorithm \cite[]{Roberts:1998,Roberts:2001}.

A simpler algorithm is the unadjusted Langevin algorithm, also known as ULA \cite[]{Parisi:1981,Ermak:1975}, which simulates from the Euler approximation but does not use a Metropolis accept-reject step and so the MCMC output produces a biased approximation of $\pi$. Computationally, such an algorithm is quicker per-iteration, but often this saving is small, as the $O(N)$ cost of calculating $\nabla U(\st)$, which is required for one step of the Euler approximation, is often at least as expensive as the cost of the accept-reject step. Furthermore, the optimal step size for MALA is generally large, resulting in a poor Euler approximation to the Langevin dynamics -- and so ULA requires a smaller step size, and potentially many more iterations.

The computational bottleneck for ULA is in calculating $\nabla U(\st)$, particularly if we have a large sample size, $N$, as $U(\st)=\sum_{i=1}^N U_i(\st)$. A solution to this problem is to use stochastic gradient Langevin dynamics \cite[SGLD][]{Welling:2011}, which avoids calculating $\nabla U(\st)$, and instead uses an unbiased estimate of it at each iteration. It is trivial to obtain an unbiased estimate using a random subsample of the terms in the sum. The simplest implementation is to choose $n<<N$ and estimate $\nabla U(\st)$ with
\begin{equation} \label{eq:U-hat-simple}
\hat{\nabla} U(\st)^{(n)} = \frac{N}{n} \sum_{i \in \mathcal{S}_n} \nabla U_i(\st),
\end{equation}
where $\mathcal{S}_n$ is a random sample, without replacement, from $\{1,\ldots,N\}$. We call this the simple estimator of the gradients, and use the superscript $(n)$ to denote the subsample size used in constructing our estimator. The resulting SGLD is given in Algorithm \ref{alg:SGLD}, and allows for the setting where the step size of the Euler discretisation depends on iteration number. \cite{Welling:2011} justified the SGLD algorithm by giving an informal argument that if the step size decreases to 0 with iteration number, then it will converge to the true Langevin dynamics, and hence be exact; see Section \ref{sec:theory} for a formal justification of this. 

\begin{algorithm}
    \caption{SGLD}
    \KwIn{$\st_0$, $\{h_0,\ldots,h_K\}$.}
    \For{$k \in 1, \dots, K$} {
        Draw $\mathcal{S}_n \subset \{1,\ldots,N\}$ without replacement \\
        Estimate $\hat{\nabla} U(\st)^{(n)}$ using \eqref{eq:U-hat-simple} \\
        Draw $\xi_k \sim N( 0, h_k I )$ \\
        Update $\st_{k+1} \leftarrow \st_k - \frac{h_k}{2} \hat{\nabla} U(\st_k)^{(n)}  + \xi_k$
    }
    \label{alg:SGLD}
\end{algorithm}

The advantage of SGLD is that, if $n<<N$, the per-iteration cost of the algorithm can be greatly smaller than either MALA or ULA. For large data applications, SGLD has been empirically shown to perform better than standard MCMC when there is a fixed computational budget \citep{ahn2015large,li2016scalable}.  In challenging examples, performance has been based on measures of prediction accuracy for a hold-out sample, rather than based on how accurately the samples approximate the true posterior. Furthermore, the conclusions from such studies will clearly depend on the computational budget, with larger budgets favouring exact methods such as MALA -- see the theoretical results in Section \ref{sec:theory}.

The SGLD algorithm is closely related to stochastic gradient descent (SGD) \citep{robbins1951stochastic}, an efficient algorithm for finding local maxima of a function. The only difference is the inclusion of the additive Gaussian noise at each iteration of SGLD. Without the noise, but with a suitably decreasing step size, stochastic gradient descent would converge to a local maxima of the density $\pi(\st)$. Again, SGLD has been shown empirically to out-perform stochastic gradient descent  \citep{chen2014stochastic} at least in terms of prediction accuracy -- intuitively this is because SGLD will give samples around the estimate obtained by stochastic gradient descent and thus can average over the uncertainty in the parameters. This strong link between SGLD and stochastic gradient descent may also explain why the former performs well when compared to exact MCMC methods, in terms of prediction accuracy.

%\CN{I think it's worth keeping this in. I think Michael Jordan has a recent paper which compares SGD against ULA and essentially shows that in the convex setting, you should use SGD, otherwise use ULA}

\subsection{Estimating the Gradient}

A key part of SGLD is replacing the true gradient with an estimate. The more accurate this estimator is, the better we would expect SGLD to perform, and thus it is natural to consider alternatives to the simple estimator (\ref{eq:U-hat-simple}). 

One way of reducing the variance of a Monte Carlo estimator is to use control variates \cite[]{Ripley:1987}, which in our setting involves choosing a set of simple functions $u_i$, $i=1,\ldots,N$, whose sum $\sum_{i=1}^N u_i(\st)$ is known for any $\st$. As
\[
 \sum_{i=1}^N \nabla U_i(\st) = \sum_{i=1}^N u_i(\st) + \sum_{i=1}^N \left(\nabla U_i(\st) - u_i(\st)  \right), 
\]
we can obtain the unbiased estimator $ \sum_{i=1}^N u_i(\st)+(N/n) \sum_{i \in \mathcal{S}_n} (\nabla U_i(\st) - u_i(\st) )$, where again $\mathcal{S}_n$ is a random sample, without replacement, from $\{1,\ldots,N\}$. The intuition behind this idea is that if each $u_i(\st)\approx \nabla U_i(\st)$, then this estimator can have a much smaller variance.

Recent works, for example \cite{Baker:2017} and \cite{Huggins:2016} \cite[see][for similar ideas used in different Monte Carlo procedures]{Bardenet:2017,Pollock:2016,Bierkens:2019}, have implemented this control variate technique with each $u_i(\st)$ set as a constant. These approaches propose (i) using stochastic gradient descent to find an approximation to the mode of the distribution we are sampling from, which we denote as $\hat{\st}$; and (ii) set $u_i(\st)=\nabla U_i(\hat{\st})$. This leads to the following control variate estimator,
\[
 \hat{\nabla}_{cv} U(\st)^{(n)}= \sum_{i=1}^N \nabla U_i(\hat{\st}) + \frac{N}{n} \sum_{i \in \mathcal{S}_n} \left( \nabla U_i(\st)-\nabla U_i(\hat{\st}) \right).
\]
Implementing such an estimator involves an up-front of cost of finding a suitable $\hat{\st}$ and calculating, storing and summing $\nabla U_i(\hat{\st})$ for $i=1,\ldots,N$. Of these, the main cost is finding a suitable $\hat{\st}$. % -- though under the strong-convexity assumptions given in Section \ref{sec:theory}, requires one pass through the full data set. 
Though we can then use $\hat{\st}$ as a starting value for the SGLD algorithm, replacing $\st_0$ with $\hat{\st}$ in Algorithm \ref{alg:SGLD}, which can significantly reduce the burn-in phase (see Figure \ref{fig:lregDim} for an illustration). 

The advantage of using this estimator can be seen if we compare bounds on the variance of this and the simple estimator. To simplify the exposition, assume that $\nabla U_i(\st)$ and its derivatives are bounded for all $i$ and $\st$. Then, under strong convexity assumptions \eqref{eq:assumptions}, there are constants $C_1$ and $C_2$ such that
\[
 \Var { \hat{\nabla} U(\st)^{(n)} } \leq C_1 \frac{N^2}{n}, ~~~ 
 \Var { \hat{\nabla}_{cv} U(\st)^{(n)} } \leq C_2 ||\st-\hat{\st}||^2 \frac{N^2}{n},
\]
where $||\cdot||$ denotes Euclidean distance. Thus, when $\st$ is close to $\hat{\st}$, we would expect the latter variance to be smaller. Furthermore, in many settings when $N$ is large we would expect a value of $\st$ drawn from the target to be of distance $O(N^{-1/2})$, thus using control variates will reduce the variance from $O(N^2/n)$ to $O(N/n)$. This simple argument suggests that, for the same level of accuracy, we can reduce the computational cost of SGLD by $O(N)$ if we use control variates. This is supported by a number of theoretical results \cite[e.g.][]{Nagapetyan:2017,Baker:2017,Brosse:2018} which show that, if we ignore the pre-processing cost of finding $\hat{\st}$, the computational cost per-effective sample size of SGLD with control variates has a computational cost that is $O(1)$, rather than the $O(N)$ for SGLD with the simple gradient estimator \eqref{eq:U-hat-simple}.

A further consequence of these bounds on the variance is that they suggest that if $\st$ is far from $\hat{\st}$ then the variance of using control variates can be larger, potentially substantially larger, than that of the simple estimator. Two ways have been suggested to deal with this. One is to only use the control variate estimator when $\st$ is close enough to $\hat{\st}$ \cite[]{Fearnhead:2018}, though it is up to the user to define what ``close enough'' is in practice. The second is to update $\hat{\st}$ during SGLD. This can be done efficiently by using $u_i(\st)=\nabla U_i(\st_{k_i})$, where $\st_{k_i}$ is the value of $\st$ at the most recent iteration of the SGLD algorithm where $\nabla U_i(\st)$ was evaluated \cite[]{Dubey:2016}.  This involves updating the storage of $u_i(\st)$ and its sum at each iteration; importantly the latter can be done with an $O(n)$ calculation. A further possibility, which we are not aware has yet been tried, is to use $u_i(\st)$ that are non-constant, and thus try to accurately estimate $\nabla U_i(\st)$ for a wide range of $\st$ values.

Another possibility for reducing the variance of the estimate of $\nabla U(\st)$ is to use preferential sampling. If we generate a sample, $\mathcal{S}_n$, such that the expected number of times $i$ appears is $w_i$, then we could use the unbiased estimator
\[
 \hat{\nabla}_{w} U(\st)^{(n)} = \sum_{i \in \mathcal{S}_n} \frac{\nabla U_i(\st)}{w_i}.
\]
The simple estimator (\ref{eq:U-hat-simple}) is a special case of this estimator where $w_i=n/N$ for all $i$. This weighted estimator can have a lower variance if we choose larger $w_i$ for $\nabla U_i(\st)$ values that are further from the mean value. A natural situation where such an estimator would make sense would be if we have data from a small number of cases and many more controls, where giving larger weights to the cases is likely to reduce the variance. Similarly, if we have observations that vary in their information about the parameters, then giving larger weights to more informative observations would make sense. Note that using weighted sampling can be combined with the control variate estimator -- with a natural choice of weights that are increasing with the size of the derivative of $\nabla U_i(\st)$ at $\hat{\st}$. We can also use stratified sampling ideas, which try to ensure each subsample is representative of the full data \citep{sen2019efficient}.

Regardless of the choice of gradient estimator, an important question is how large should the subsample size be? A simple intuitive rule, which has some theoretical support \cite[e.g.][]{Vollmer:2016,Nagapetyan:2017}, is to choose the subsample size such that if we consider one iteration of SGLD, the variance of the noise from the gradient term is dominated by the variance of the injected noise. As the former scales like $h^2$ and the latter like $h$ then this suggests that as we reduce the step size, $h$, smaller subsample sizes could be used -- see Section \ref{sec:theory-example} for more details.

\subsection{Theory for SGLD} \label{sec:theory}

As described so far, SGLD is a simple and computationally efficient approach to approximately sample from a stochastic process whose asymptotic distribution is $\pi$; but how well do samples from SGLD actually approximate $\pi$? In particular, whilst for small step sizes the approximation within one iteration of SGLD may be good, do the errors from these approximations accumulate over many iterations? There is now a body of theory addressing these questions. Here we give a brief, and informal overview of this theory. We stress that all results assume a range of technical conditions on $\pi(\st)$, some of which are strong -- see the original references for details. In particular, most results assume that the drift of the underlying Langevin diffusion will push $\st$ towards the centre of the distribution, an assumption which means that the underlying Langevin diffusion will be geometrically ergodic, and an assumption that is key to avoid the accumulation of error within SGLD.

%SOMETHING MORE ON THIS?

There are various ways of measuring accuracy of SGLD, but current theory focuses on two approaches. The first considers estimating the expectation of a suitable test function $\phi(\st)$, i.e. $\Expects{\pi}{\phi(\st)}=\int \pi(\st)\phi(\st) \mbox{d}\st$, using an average over the output from $K$ iterations of SGLD, $(1/K)\sum_{k=1}^K \phi(\st_k)$. In this setting, we can measure the accuracy of the SGLD algorithm through the mean square error of this estimator. \cite{Teh:2016} consider this in the case where the SGLD step size $h_k$ decreases with $k$. %, as originally suggested by \cite{Welling:2011}. 
The mean square error of the estimator can be partitioned into a square bias term and a variance term. For large $K$, the bias term increases with the step size, whereas the variance term is decreasing. \cite{Teh:2016} show that in terms of minimising the asymptotic mean square error, the optimal choice of step size should decrease as $k^{-1/3}$, with the resulting mean square error of the estimator decaying as $K^{-2/3}$. This is slower than for standard Monte Carlo procedures, where a Monte Carlo average based on $K$ samples will have mean square error that decays as $K^{-1}$. The slower rate comes from needing to control the bias as well as the variance, and is similar to rates seen for other Monte Carlo problems where there are biases that need to be controlled \cite[e.g. Section 3.3 of][]{Fearnhead:2008}. In practice, SGLD is often implemented with a fixed step size $h$. \cite{Vollmer:2016} give similar results on the bias-variance trade-off for SGLD with a fixed step size, with a mean square error for $K$ iterations and a step size of $h$ being $O(h^2+1/(hK))$. The $h^2$ term comes from the squared bias and $1/hK$ from the variance term.  The rate-optimal choice of $h$ as a function of $K$ is $K^{-1/3}$, which again gives an asymptotic mean square error that is $O(K^{-2/3})$;  the same asymptotic rate as for the decreasing step size. This result also shows that with larger computational budgets we should use smaller step sizes. Furthermore, if we have a large enough computational resource then we should prefer exact MCMC methods over SGLD: as computing budget increases, exact MCMC methods will eventually be more accurate.

The second class of results consider the distribution that SGLD samples from at iteration $K$ with a given initial distribution and step size. Denoting the density of $\st_K$ by $\tilde\pi_K(\st)$, one can the measure an appropriate distance between $\tilde\pi_K(\st)$ and $\pi(\st)$. The most common distance used is the Wasserstein distance \cite[]{Gibbs:2002}, primarily because it is particularly amenable to analysis. %the error between the dynamics of the Langevin diffusion and SGLD. 
Care must be taken when interpreting the Wasserstein distance, as it is not scale invariant -- so changing the units of our parameters will result in a corresponding scaling of the Wasserstein distance between the true posterior and the approximation we sample from. Furthermore, as we increase the dimension of the parameters, $d$, and maintain the same accuracy for the marginal posterior of each component, the Wasserstein distance will scale like $d^{1/2}$.

There are a series of results for both ULA and SGLD in the literature \cite[]{Dalalyan:2017B,Dalalyan:2017,Durmus:2017,Chatterji:2018,Brosse:2018}. Most of this theory assumes strong-convexity of the log-target density \cite[see][for similar theory under different assumptions]{raginsky2017non,majka2018non}, which means that there exists strictly positive constants, $0<m\leq M$, such that for all $\st$, and $\st'$,
\begin{equation}
  \label{eq:assumptions}
 ||\nabla U(\st)-\nabla U(\st')||_2 \leq M ||\st-\st'||_2, ~~\text{and}~~ U(\st)-U(\st')-\nabla U(\st')^\top(\st-\st')\geq \frac{m}{2}||\st-\st'||_2^2,  
\end{equation}
where $||\cdot||_2$ denotes the Euclidean norm. If $U(\st)$ is twice continuously differentiable, these conditions are equivalent to assuming upper and lower bounds on all possible directional derivatives of $U(\st)$. The first bound governs how much the drift of the Langevin diffusion can change, and is important in the theory for specifying appropriate step-lengths, which should be less than $1/M$, to avoid instability of the Euler discretisation; it also ensures that the target density is uni-modal. The second bound ensures that the drift of the Langevin will push $\st$ towards the centre of the distribution, an assumption which means that the underlying Langevin diffusion will be geometrically ergodic, and consequently is key to avoiding the accumulation of error within SGLD.

For simplicity, we will only informally present results from \cite{Dalalyan:2017}, as these convey the main ideas in the literature. 
These show that, for $h<1/(M+m)$, the Wasserstein-2 distance between $\tilde\pi_K(\st)$ and $\pi(\st)$, denoted $\mathcal{W}_2(\tilde\pi_K,\pi)$ can be bounded as
\begin{equation} \label{eq:Wasserstein-bd1}
 \mathcal{W}_2(\tilde\pi_K,\pi) \leq (1-mh)^K \mathcal{W}_2(\tilde\pi_0,\pi) + C_1(hd)^{1/2} + C_2 \sigma (hd)^{1/2},
\end{equation}
where $m$, $C_1$ and $C_2$ are constants, $d$ is the dimension of $\st$, and $\sigma^2$ is a bound on the variance of the estimate for the gradient. Setting $\sigma^2=0$ gives a Wasserstein bound for the ULA approximation. The first term on the right-hand side measures the bias due to starting the SGLD algorithm from a distribution that is not $\pi$, and is akin to the bias due to finite burn-in of the MCMC chain. Providing $h$ is small enough, this will decay exponentially with $K$. The other two terms are, respectively, the effects of the approximations from using an Euler discretisation of the Langevin diffusion and an unbiased estimate of $\nabla U(\st)$.  
%If we use the simple estimator of $\nabla U(\st)$ \eqref{eq:U-hat-simple}, whose variance if $O(N^2/n)$, the latter term is $O(N\sqrt{hd/n}). dominates the other terms in the Wasserstein bound. A corollary of these results is that if we assume $M$ and $m$ scale linearly with $N$, then for a target level of precision  $\epsilon$ to achieve $ \mathcal{W}_2(\tilde\pi_K,\pi) \leq \epsilon$ requires $O(d/n\epsilon^2)$ iterations of the SGLD algorithm (Alg. \ref{alg:SGLD}). Using the control variate gradient estimator $ \hat{\nabla}_{cv} U(\st)^{(n)}$ gives an improved $O(d/N\epsilon^2)$ mixing time \citep[see][for results pertaining to other stochastic gradient MCMC algorithms]{Chatterji:2018}. 

A natural question is, what do we learn from results such as (\ref{eq:Wasserstein-bd1})? These results give theoretical justification for using SGLD, and show we can sample from an arbitrarily good approximation to our posterior distribution if we choose $K$ large enough, and $h$ small enough. They have also been used to show the benefits of using control variates when estimating the gradient, which results in a computational cost that is $O(1)$, rather than $O(N)$, per effective sample size \cite{Baker:2017,Chatterji:2018}. Perhaps the main benefit of results such as (\ref{eq:Wasserstein-bd1}) is that they enable us to compare the properties of the different variants of SGLD that we will introduce in Section \ref{sec:altern-stoch-grad}, and in particular how different algorithms scale with dimension, $d$ (see Section \ref{sec:altern-stoch-grad} for details). However, they do not directly tell us how to choose $K$, $h$ or the subsample size in practice.

Perhaps more importantly than having a quantitative measure of approximation error is to have an idea as to the form of the error that the approximations in SGLD induce. Results from \cite{Vollmer:2016} and \cite{Brosse:2018}, either for specific examples or for the limiting case of large $N$, give insights into this. For an appropriately implemented SGLD algorithm, and for large data size $N$, these results show that the distribution we sample from will asymptotically have the correct mode but will inflate the variance. We discuss ways to alleviate this in the next section when we consider a specific example.

\subsection{A Gaussian Example} \label{sec:theory-example}

To gain insight into the properties of SGLD, it is helpful to consider a simple tractable example where we sample from a Gaussian target. We will consider a 2-dimensional Gaussian, with variance $\bSigma$ and, without loss of generality, mean zero. The variance matrix can be written as $\bP^{\top}\bD\bP$ for some rotation matrix $\bP$ and diagonal matrix $\bD$, whose entries satisfy the condition $\sigma_1^2\geq \sigma_2^2$. For this model, the drift term of the Langevin diffusion is
\[
\nabla U(\st)=- \bSigma^{-1}\st = -\bP^{\top}\bD^{-1}\bP\st.
\]
The $k$th iteration of the SGLD algorithm is
\begin{equation}
  \label{eq:gaussian-example-dynamics}
 \st_{k}= \st_{k-1}+\frac{h}{2}\hat{\nabla} U(\st_{k-1})+\sqrt{h}\bZ= \st_{k-1}-\frac{h}{2}\bP^{\top}\bD^{-1}\bP\st_{k-1}+h \bnu_k + \sqrt{h}\bZ_k,
\end{equation}
where $\bZ_k$ is a vector of two independent standard normal random variables and $\bnu_k$ is the error in our estimate of $\nabla U(\st_{k-1})$. The entries of $\bD^{-1}$ correspond to the constants that appear in condition (\ref{eq:assumptions}), with $m=1/\sigma_1^2$ and $M=1/\sigma_2^2$.

To simplify the exposition, it is helpful to study the SGLD algorithm for the transformed state $\tilde{\st}=\bP\st$, for which we have
\[
 \tilde{\st}_k=\tilde{\st}_{k-1}-\frac{h}{2}\bD^{-1}\tilde{\st}_{k-1}+h\bP\bnu + \sqrt{h}\bP\bZ=\left(\begin{array}{cc}1-h/(2\sigma_1^2) & 0 \\ 0 & 1-h/(2\sigma_2^2)  \end{array} \right)\tilde{\st}_{k-1}+h\bP\bnu_k + \sqrt{h}\bP\bZ_k.
\]
As $\bP$ is a rotation matrix, the variance of $\bP\bZ_k$ is still the identity.

In this case, the SGLD update is a vector auto-regressive process. This process will have a stationary distribution provided $h<4\sigma_2^2=4/M$, otherwise the process will have trajectories that will go to infinity in at least one component. 
This links to the requirement of a bound on the step size that is required in the theory for convex target distributions described above.

Now assume $h<2\sigma_2^2$, and write $\lambda_j=h/(2\sigma_j^2)<1$. We have the following dynamics for each component, $j=1,2$
\begin{equation}
  \label{eq:1d-example}
  \tilde{\st}_k^{(j)}=(1-\lambda_j)^k \tilde{\st}_0^{(j)}+\sum_{i=1}^k (1-\lambda_j )^{k-i}\left(h\bP\bnu_i^{(j)} + \sqrt{h}\bP\bZ_i^{(j)} \right),  
\end{equation}
where $\tilde{\st}_k^{(j)}$ is the $j$th component of $\tilde{\st}_k$, and similar notation is used for $\bnu_i$ and $\bZ_i$. From this, we immediately see that SGLD forgets its initial condition exponentially quickly. However, the rate of exponential decay is slower for the component with larger marginal variance, $\sigma_1^2$. Furthermore, as the size of $h$ is constrained by the smaller marginal variance $\sigma_2^2$, this rate will necessarily be slow if $\sigma_2^2<< \sigma_1^2$; this suggests that there are benefits of re-scaling the target so that marginal variances of different components are roughly equal.

Taking the expectation of \eqref{eq:1d-example}  with respect to $\bnu$ and $\bZ$, and letting $k \rightarrow \infty$, results in SGLD dynamics that have the correct limiting mean but with an inflated variance. This is most easily seen if we assume that the variance of $\bP\bnu$ has a variance that is independent of position, $\bV$ say. In this case, the stationary distribution of SGLD will have variance 
\[
\Vars{\tilde{\pi}}{\tilde{\st}}= \left(\begin{array}{cc} (1-(1-\lambda_1)^2 )^{-1} & 0 \\ 0 &  (1-(1-\lambda_2)^2)^{-1}\end{array} \right) (h^2\bV+h\bI),
\]
where $\bI$ is the identity matrix. The marginal variance for component $j$ is thus 
\[
\sigma_j^2\frac{1+h\bV_{jj}}{1-h/(4\sigma_j^2)} = \sigma_j^2\left(1+h\bV_{jj}\right)+\frac{h}{4}+O(h^2).
\]
The inflation in variance comes both from the noise in the estimate of $\nabla U(\st)$, which is the $h\bV_{jj}$ factor, and the Euler approximation, through the additive constant, $h/4$. 
For more general target distributions, the mean of the stationary distribution of SGLD will not necessarily be correct, but we would expect the mean to be more accurate than the variance, with the variance of SGLD being greater than that of the true target.
The above analysis further suggests that, for targets that are close to Gaussian, it may be possible to perform a better correction to compensate for the inflation of  the variance. \cite{Vollmer:2016} suggest reducing the driving Brownian noise \cite[see also][]{chen2014stochastic}. That is, we replace $\bZ_k$ by Gaussian random variables with a covariance matrix so that the covariance matrix of  $h\bnu_k+\sqrt{h}\bZ$ is the identity. If the variance of $\bnu_k$ is known, then \cite{Vollmer:2016} show that this can substantially improve the accuracy of SGLD. In practice, however, it is necessary to estimate this variance and it is an open problem as to how one can estimate this accurately enough to make the idea work well in practice \cite[]{Vollmer:2016}

%\PF{Perhaps look at MSE of estimate?? or show a plot of coupled Langevin and SGLD dynamics...}

%\PF{Perhaps I should add comments that similar analysis would apply to $d$ dimensional Gaussian, with the key criteria being $h$ which depends on the smallest variance relative to the other marginal variances.}

\section{A General Framework for Stochastic Gradient MCMC}
\label{sec:altern-stoch-grad}

%\PF{Why the switch from $\st$ to $\bx$?} \CN{it's not ideal, but it seemed like the best way to introduce auxiliary variables.}

%\CN{CHRIS TO DO: THE TIDIEST WAY TO DO THIS WOULD PROBABLY BE TO PRESENT THE COMPLETE RECIPE APPROACH, WHICH GIVES SGLD, SG-HMC, ETC. AS SPECIAL CASES.} \todo{the complete recipe and specific examples such as SG-HMC; Jack's CIR-sampler/pointing out the potential for using other diffusions you can simulate exactly from; summary of theory}

%The Langevin diffusion provides the basis for gradient-based MCMC algorithms like ULA and MALA, which are based on Langevin dynamics. 

So far we have considered stochastic gradient MCMC based on approximating the dynamics of the Langevin diffusion. However, we can write down other diffusion processes that have $\pi$ as their stationary distribution, and use similar ideas to approximately simulate from one of these. A general approach to doing this was suggested by \cite{ma2015complete} and leads to a much wider class of stochastic gradient MCMC algorithms, including stochastic gradient versions of popular MCMC algorithms such as Hamiltonian Monte Carlo \citep{Neal:2011,Carpenter:2017}. 

The class of diffusions we will consider may include a set of auxiliary variables. As such, we let $\bzeta$ be a general state, with the assumption that this state contains $\st$. For example, for the Langevin diffusion $\bzeta=\st$; but we could mimic Hamiltonian MCMC and introduce an auxiliary velocity component, $\brho$, in which case $\bzeta=(\st,\brho)$.  We start by considering a general stochastic differential equation for $\bzeta$,
\begin{equation}
  \label{eq:sde}
\mbox{d}\bzeta= \frac{1}{2}\mathbf{b}(\bzeta)\mbox{d}t+\sqrt{\bD(\bzeta)}\mbox{d}B_t,
\end{equation}
where the vector $\mathbf{b}(\bzeta)$ is the drift component, $\bD(\bzeta)$ is a positive semi-definite diffusion matrix, and $\sqrt{\bD(\bzeta)}$ is any square-root of $\bD(\bzeta)$. \cite{ma2015complete} show how to choose $\mathbf{b}(\bzeta)$ and $\bD(\bzeta)$ such that (\ref{eq:sde}) has a specific stationary distribution. 
We define the function $H(\bzeta)$ such that  $\exp\{-H(\bzeta)\}$ is intergrable and let $\bQ(\bzeta)$ be a skew-symmetric curl matrix, so $\bQ^\top=-\bQ$. Then the choice
\begin{equation}
  \label{eq:sde-f}
  \mathbf{b}(\bzeta) = -\left[\bD(\bzeta)+\bQ(\bzeta)\right]\nabla H(\bzeta) + \Gamma(\bzeta) ~~\text{and}~~ \Gamma_i(\bzeta) = \sum_{j=1}^d \frac{\partial}{\partial \bzeta_{j}}(\bD_{ij}(\bzeta)+\bQ_{ij}(\bzeta)),
\end{equation}
ensures that the stationary distribution of (\ref{eq:sde}) is proportional to $\exp\{-H(\bzeta)\}$. \cite{ma2015complete} show that any diffusion process with a stationary distribution proportional to $\exp\{-H(\bzeta)\}$ is of the form (\ref{eq:sde}) with the drift and diffusion matrix satisfying (\ref{eq:sde-f}). To approximately sample from our diffusion, we can employ the same discretisation of the continuous-time dynamics that we used for the Langevin diffusion \eqref{eq:Euler},
\begin{equation}
  \label{eq:complete-recipe-approx}
      \bzeta_{t+h}\approx \bzeta_t - \frac{h}{2} \left[(\bD(\bzeta_t)+\bQ(\bzeta_t))\nabla H(\bzeta_t)+\Gamma(\bzeta_t)\right] + \sqrt{h} \bZ, \quad t \geq 0,
\end{equation}
where $\bZ \sim N(0,\bD(\bzeta_t))$. The diffusions we are interested in have a stationary distribution where the $\st$-marginal distribution is $\pi$. If $\bzeta=\st$ then this requires $H(\bzeta)=U(\st)$. If, however, $\bzeta$ also includes some auxiliary variables, say $\brho$, then this is most easily satisfied by setting $H(\bzeta)=U(\st)+K(\brho)$ for some suitable function $K(\brho)$. This choice leads to a stationary distribution under which $\st$ and $\brho$ are independent.

We can derive a general class of \textit{stochastic gradient MCMC} (SGMCMC) algorithms, where we simply replace the gradient estimate $\nabla H(\bzeta_t)$ with an unbiased estimate $\hat \nabla H(\bzeta_t)$, based on data subsampling. \cite{ma2015complete} suggest that one should also correct for the variance of the estimate of the gradient, as illustrated in the example from Section \ref{sec:theory-example}, to avoid the inflation of variance in the approximate target distribution. If the variance of our estimator $\hat \nabla H(\bzeta_t)$ is ${\bV}(\st_t)$, then this inflates the conditional variance of $\bzeta_{t+h}$ given $\bzeta_t$ in (\ref{eq:complete-recipe-approx}) by $h^2\mathbf{B}(\bzeta_t)$ where
\[
\mathbf{B}(\bzeta_t)=\frac{1}{4}(\bD(\bzeta_t)+\bQ(\bzeta_t)) \bV(\st_t) (\bD(\bzeta_t)+\bQ(\bzeta_t))^\top.
\]
Given an estimate $\hat{\mathbf{B}}(\bzeta_t)$, we can correct for the inflated variance by simulating $\bZ\sim N(0,\bD(\bzeta_t)-h\hat{\mathbf{B}}(\bzeta_t) )$. Obviously, this requires that $\bD(\bzeta_t)-h\hat{\mathbf{B}}(\bzeta_t)$ is positive semi-definite. In many cases this can be enforced if $h$ is small enough. If this is not possible, then that suggests the resulting SGMCMC algorithm will be unstable; see below for an example.

The diffusion $\bD(\bzeta)$ and curl $\bQ(\bzeta)$ matrices can take various forms and the choice of matrices will affect the rate of convergence of the MCMC samplers. The diffusion matrix $\bD(\bzeta)$ controls the level of noise introduced into the dynamics of \eqref{eq:complete-recipe-approx}. When $||\bD(\bzeta)||$ is large, there is a greater chance that the sampler can escape local modes of the target, and setting $||\bD(\bzeta)||$ to be small increases the accuracy of the sampler within a local mode. Between modes of the target, the remainder of the parameter space is represented by regions of low probability mass where we would want our MCMC sampler to quickly pass through. The curl matrix $\bQ(\bzeta)$ controls the sampler's non-reversible dynamics which allows the sampler to quickly traverse low-probability regions, this is particularly efficient when the curl matrix adapts to the geometry of the target.

In Table \ref{tab:recipes} we define $H(\bzeta)$, $\bD(\bzeta)$ and $\bQ(\bzeta)$ for several gradient-based MCMC algorithms. The two most common are SGLD, which we introduced in the previous section, and SG-HMC \citep{chen2014stochastic}. This latter process introduces a velocity component that can help improve mixing, as is seen in more standard Hamiltonian MCMC methods. The closest link with the dynamics used in Hamiltonian MCMC is when $\bD(\bzeta)$ is set to be the zero-matrix. However \citep{chen2014stochastic} show that this leads to an unstable process that diverges as a result of the accumulation of noise in the estimate of the gradient; a property linked to the fact that $\bD(\bzeta)-h\hat{\mathbf{B}}(\bzeta)$ is not positive semi-definite for any $h$. The choice of $\bD(\bzeta)$ given in Table \ref{tab:recipes} avoids this problem, with the resulting stochastic differential equation being the so-called under-damped Langevin diffusion.

As discussed in Section \ref{sec:theory-example} with regard to SGLD, re-parameterising the target distribution so that the components of $\st$ are roughly uncorrelated and have similar marginal variances, can improve mixing. An extension of this idea is to adapt the dynamics locally to the curvature of the target distribution -- and this is the idea behind Riemannian versions of SGLD and SG-HMC, denoted by SG-RLD \citep{patterson2013stochastic} and SG-RHMC \citep{ma2015complete} in Table \ref{tab:recipes}. The challenge with implementing either of these algorithms is obtaining an accurate, yet easy to compute, estimate of the local curvature. A simpler approach is the stochastic gradient Nose-Hoover thermostat (SG-NHT)  \citep{ding2014bayesian} algorithm, which introduces state dependence into the curl matrix. This can be viewed as an extension of SG-HMC which adaptively controls for the excess noise in the gradients. Obviously, there are many other algorithms that could be derived from this general framework.

\begin{table}
   %\label{tab:recipes}
  \caption{A list of popular SGMCMC algorithms highlighting how they fit within the general stochastic differential equation framework \eqref{eq:sde}-\eqref{eq:sde-f}. Most of the terms are defined in the text, except: $\mathbf{C} \succeq h\bV(\st)$, which is a positive semi-definite matrix; $G(\st)$ is the Fisher information metric; $A$ is a tuning parameter for SG-NHT.}
  \label{tab:recipes} 
  \begin{tabular}{|c | c | c | c | c|}
 \hline
 Algorithm & $\bzeta$ & $H(\bzeta)$ & $\bD(\bzeta)$ & $\bQ(\bzeta)$ \\ [0.5ex] 
 \hline\hline
 SGLD & $\st$ & $U(\st)$ & $\mathbf{I}$ & $\mathbf{0}$ \\ 
 \hline
 SG-RLD &$\st$ & $U(\st) $ & $G(\st)^{-1}$ & $\mathbf{0}$ \\
 \hline
SG-HMC & $(\st,\brho)$ & $U(\st) + \frac{1}{2}\brho^\top\brho$ & $\left( \begin{array}{cc} 0 & 0 \\ 0 & \mathbf{C} \end{array}\right)$ & $\left( \begin{array}{cc} 0 & -\mathbf{I} \\ \mathbf{I} & 0 \end{array}\right)$\\
 \hline
 SG-RHMC & $(\st,\brho)$ & $U(\st) + \frac{1}{2}\brho^\top\brho$ & $\left( \begin{array}{cc} 0 & 0 \\ 0 & G(\st)^{-1} \end{array}\right)$ &  $\left( \begin{array}{cc} 0 & -G(\st)^{-1/2} \\ G(\st)^{-1/2} & 0 \end{array}\right)$ \\
 \hline
    SG-NHT & $(\st,\brho, \eta)$ & $\begin{array}{c} U(\st)+ \frac{1}{2}\brho^\top\brho \\ +\frac{1}{2d}(\eta-A)^2 \end{array}$ & $\left( \begin{array}{ccc} 0 & 0 & 0 \\ 0& A \cdot \mathbf{I} & 0 \\ 0 & 0 & 0 \end{array}\right)$ & $\left( \begin{array}{ccc} 0 & -\mathbf{I} & 0 \\ \mathbf{I} & 0 & \brho^\top/d \\ 0 & -\brho^\top/d & 0 \end{array}\right)$ \\
 \hline
\end{tabular}
\end{table}

%It can be seen from Table \ref{tab:recipes} that the complete recipe scheme \eqref{eq:complete-recipe-approx} generalises many popular gradient-based MCMC algorithms and that many simpler algorithms can be viewed as special cases of more complex algorithms. TO COMPLETE..... In Section \ref{sec:simulation-study}, we provide a numerical comparison of some of the algorithms given in Table \ref{tab:recipes}.

\subsection{Theory for SG-HMC}
\label{sec:theory-sg-hmc}

It is natural to ask which of the algorithms presented in Table \ref{tab:recipes} is most accurate. We will study this question empirically in Section \ref{sec:simulation-study}, but here we briefly present some theoretical results that compare SG-HMC with SGLD for smooth and strongly log-concave target densities. These results are for bounds on the Wasserstein distance between the target distribution and the distribution of the SGMCMC algorithm samples at iteration $k$, for an optimally chosen step size \citep{Cheng:2017}. %We presented such results for SGLD in Section \ref{sec:theory}, and similar results for SG-HMC are given in \cite{Cheng:2017}.
 The simplest comparison of the efficiencies of the two algorithm is for the case where the gradients are estimated without error. For a given level of accuracy, $\epsilon$, measured in terms of Wasserstein distance, SGLD requires $O(d^2/\epsilon^2)$ iterations, whereas SG-HMC requires $O(d/\epsilon)$ iterations. This suggests that SG-HMC is to be preferred, and the benefits of SG-HMC will be greater in higher dimensions. Similar results are obtained when using noisy estimates of the gradients, providing the variance of the estimates is small enough. However, \cite{Cheng:2017} show that there is a phase-transition in the behaviour of SG-HMC as the variance of the gradient estimates increases: if it is too large, the SG-HMC behaves like SGLD and needs a similar order of iterations to achieve a given level of accuracy.

\section{Diagnostic Tests}
\label{sec:diagnostic-tests}

When using an MCMC algorithm the practitioner wants to know if the algorithm has converged to the stationary distribution, and how to tune the MCMC algorithm to maximise the efficiency of the sampler. %There are range of diagnostic tests which use the posterior samples to assess convergence. These include looking at trace plots, comparing of the posterior means at the start and end of the chain \citep{geweke1991evaluating}, assessing the within-chain and between-chain variance \citep{gelman1992inference} or using hypothesis testing \citep{heidelberger1983simulation}. Additionally, measures of posterior sample quality, such as effective sample size and asymptotic variance, can be used to tune the MCMC sampler (i.e. select the step size parameter). 
 In the case of stochastic gradient MCMC, the target distribution is not the stationary distribution and therefore our posterior samples represent an asymptotically biased approximation of the posterior. Standard MCMC diagnostic tests \citep{brooks1998general} do not account for this bias and therefore are not appropriate for either assessing convergence or tuning stochastic gradient MCMC algorithms. %Therefore, alternative metrics are required for approximate MCMC algorithms which balance between bias and variance and can detect non-convergence of the MCMC algorithm, as well as be used to tune the sampler. 
 The design of appropriate diagnostic tests for stochastic gradient MCMC is a relatively new area of research, and currently methods based on Stein's discrepancy \citep{gorham2015measuring,gorham2016measuring,gorham2017measuring} are the most popular approach. These methods provide a general way of assessing how accurately a sample of values approximate a distribution.
 
Assume we have a sample, say from an SGMCMC algorithm, $\st_1,\st_2,\ldots,\st_K \in \mathbb{R}^d$, and denote the empirical distribution that this sample defines as $\tilde{\pi}$. We can define a measure of how well this sample approximates a target distribution, $\pi$, through how close expectations under $\tilde{\pi}$ are to the expectations under $\pi$. If they are close for a broad class of functions, $\mathcal{H}$, then this suggests the approximation error is small. This  motivates the following measure of discrepancy, 
\begin{equation}
  \label{eq:ipm}
  d_\mathcal{H}(\tilde{\pi},\pi) := \sup_{\hslash \in \mathcal{H}} |\Expects{\tilde{\pi}}{\hslash(\st)}-\Expects{\pi}{\hslash(\st)}|,  
\end{equation}
where $\Expects{\tilde{\pi}}{\hslash(\st)} = \frac{1}{K}\sum_{k=1}^K \hslash(\st_k)$ is an approximation of $\Expects{\pi}{\hslash(\st)}$. For appropriate choices of $\mathcal{H}$, it can be shown that if we denote the approximation from a sample of size $K$ by $\tilde{\pi}_K$, then $d_\mathcal{H}(\tilde{\pi}_K,\pi)\rightarrow 0$ if and only if  $\tilde{\pi}_K$ converges weakly to $\pi$. Moreover, even if this is not the case, if functions of interest are in $\mathcal{H}$ then a small value of $d_\mathcal{H}(\tilde{\pi},\pi) $ would mean that we can accurately estimate posterior expectations of functions of interest.

Unfortunately, \eqref{eq:ipm} is in general intractable as it depends on the unknown $\Expects{\pi}{\hslash(\st)}$. The Stein discrepancy approach circumvents this problem by using a class, $\mathcal{H}$, that only contains functions whose expectation under $\pi$ are zero.  We can construct such functions from stochastic processes, such as the Langevin diffusion, whose invariant distribution is $\pi$. If the initial distribution of such a process is chosen to be $\pi$ then the expectation of the state of the process will be constant over time. Moreover, the rate of change of expectations can be written in terms of the expectation of the generator of the process applied to the function: which means that functions that can be written in terms of the generator applied to a function will have expectation zero under $\pi$.

%The disadvantage of this approach is that we only get to indirectly specify our class of functions, $\mathcal{H}$, used to define the discrepancy. We instead choose an operator, $\mathcal{T}$ say, that is the generator of e.g. the Langevin diffusion, a class of functions $\mathcal{G}$ that the operator can act on, and then $\mathcal{H}$ is the set of functions of the form $\mathcal{T}g$ for $g \in \mathcal{G}$. Much of the work in this area \citep{gorham2015measuring,gorham2016measuring,gorham2017measuring}  has focused on how to choose $\mathcal{T}$ and $\mathcal{G}$ such that it is possible to evaluate the discrepancy \eqref{eq:ipm} and that the discrepancy will detect both convergence and non-convergence to $\pi$.

In our experience, the computationally most feasible approach, and easiest to implement, is the kernel Stein set approach of \cite{gorham2017measuring}, which enables the discrepancy to be calculated as a sum of some kernel evaluated at all pairs of points in the sample. As with all methods based on Stein discrepancies, it also requires the gradient of the target at each sample point -- though we can use unbiased noisy estimates for these \citep{gorham2017measuring}. The kernel Stein discrepancy is defined as

\begin{equation}
 \label{eq:ksd}
KSD(\tilde{\pi}_K,\pi) := \sum_{j = 1}^{d} \sqrt{\sum_{k,k' = 1}^{K} \frac{k_j^0(\st_{k}, \st_{k'})}{K^2}},
\end{equation}
where the Stein kernel for $j \in \{1,\ldots,d\}$  is given by  
\begin{align*}
 k_j^0(\st,\st^\prime) = & (\nabla_{\st^{(j)}} U(\st) \nabla_{\st^{\prime(j)}} U(\st^\prime)) k(\st,\st^\prime) + \nabla_{\st^{(j)}} U(\st) \nabla_{\st^{\prime(j)}} k(\st,\st^\prime) \\
+  & \nabla_{\st^{\prime(j)}} U(\st^\prime) \nabla_{\st^{(j)}} k(\st,\st^\prime) + \nabla_{\st^{(j)}} \nabla_{\st^{\prime(j)}} k(\st,\st^\prime). 
\end{align*}
The kernel $k$ has to be carefully chosen, particularly when $d \geq 3$, as some kernel choices, e.g. Gaussian and Matern, result in a kernel Stein discrepancy which does not detect non-convergence to the target distribution. \cite{gorham2017measuring} recommend using the inverse multi-quadratic kernel, $k(\st,\st^\prime) = (c^2 + ||\st-\st^\prime||_2^2)^\beta$,which they prove detects non-convergence when $c>0$ and $\beta \in (-1,0)$. A drawback of most Stein discrepancy measures, including the kernel Stein method, is that the computational cost scales quadratically with the sample size. This is more computationally expensive than standard MCMC metrics (e.g. effective sample size), however, the computation can be easily parallelised to give faster calculations.

We illustrate the kernel Stein discrepancy on the Gaussian target introduced in Section \ref{sec:theory-example}, where we choose diagonal and rotation matrices
\begin{align*}
  D = \left(\begin{array}{cc} 2 & 0 \\ 0 & 1  \end{array} \right) \ \ \mbox{and} \ \ P = \left(\begin{array}{cc} \cos\frac{\pi}{4} & \sin\frac{\pi}{4} \\ -\sin\frac{\pi}{4} & \cos\frac{\pi}{4}  \end{array} \right).
\end{align*}

\begin{figure}[h]
  \centering
  \includegraphics[scale=0.5]{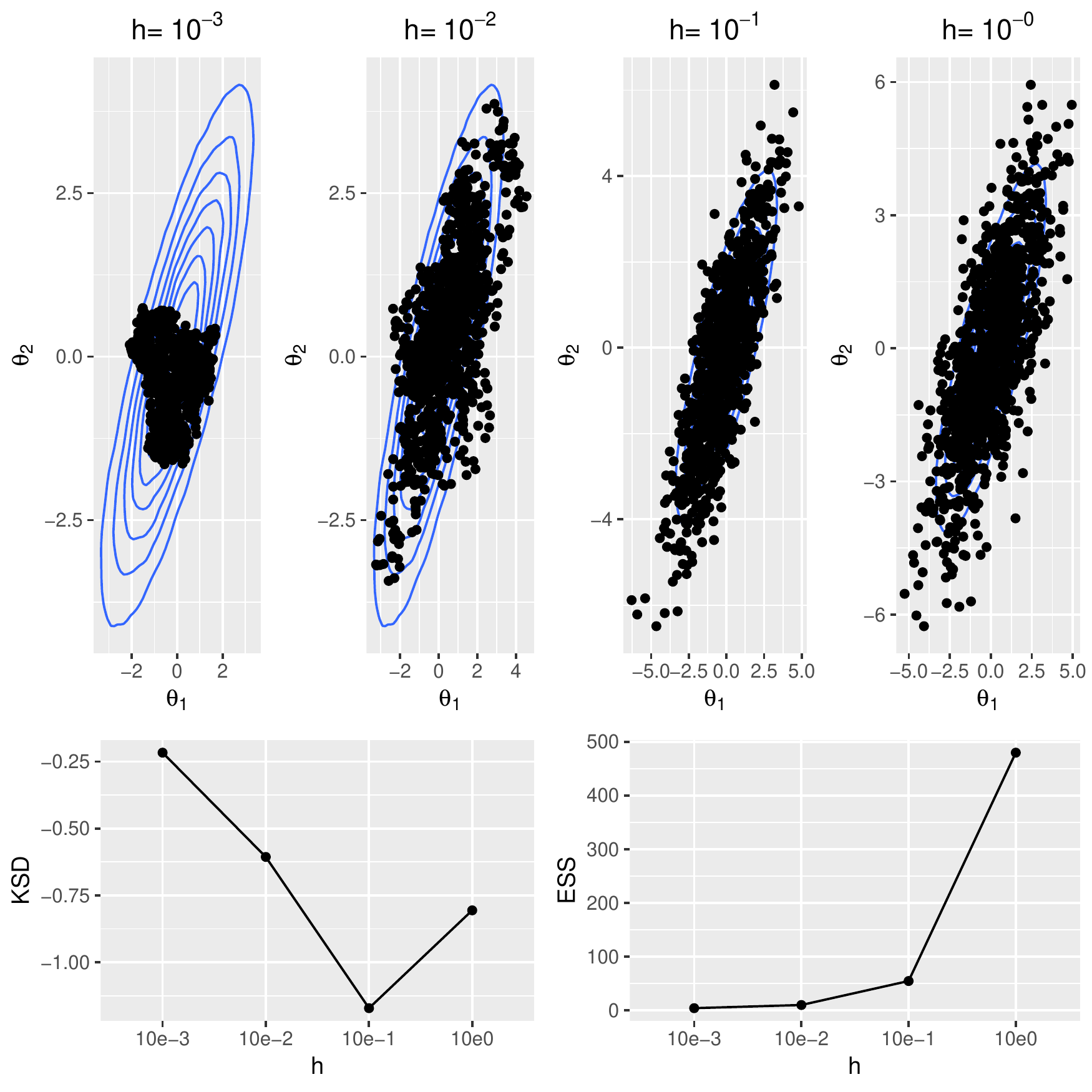}
  \caption{Top: Samples generated from the Langevin dynamics \eqref{eq:gaussian-example-dynamics} are plotted over the bivariate Gaussian target. The samples are thinned to 1,000 for the ease of visualisation. Bottom: The kernel Stein discrepancy (log10) and effective sample size are calculated for each Markov chain with varying step size parameter $h$.}
  \label{fig:ksd} 
\end{figure}

We iterate the Langevin dynamics \eqref{eq:gaussian-example-dynamics} for 10,000 iterations, starting from $\st=(0,0)$ and with noisy gradients simulated as the true gradient plus noise, $\bnu_k \sim N(0,0.01)$. We test the efficiency of the Langevin algorithm in terms of the step size parameter $h$ and use the kernel Stein discrepancy metric \eqref{eq:ksd} to select a step size parameter which produces samples that most closely approximate the target distribution. We consider a range of step size parameters $h= \{10^{-3},10^{-2},10^{-1},10^0\}$ which satisfy the requirement that $h<4\sigma_2^2$ to prevent divergent chains. In Figure \ref{fig:ksd} we plot the samples generated from the Langevin algorithm for each of the step size parameters. We also calculate the kernel Stein discrepancy \eqref{eq:ksd}and effective sample size for each Markov chain. Visually, it is clear from Figure \ref{fig:ksd} that $h=0.1$ produces samples which most closely represent the target distribution. A large value for $h$ leads to over-dispersed samples and a small $h$ prevents the sampler from exploring the whole target space within the fixed number of iterations. Setting $h=0.1$ also gives the lowest kernel Stein discrepancy, whereas $h=1$ maximises the effective sample size. This supports the view that effective sample size and other standard MCMC metrics, which do not account for sample bias, are not appropriate diagnostic tools for stochastic gradient MCMC.

\section{Extending the SGMCMC framework}
\label{sec:extend-stoch-grad}

Under the general SGMCMC framework outlined in Section \ref{sec:altern-stoch-grad}, it is possible to extend the SGLD algorithm beyond Langevin dynamics and consider a larger class of MCMC algorithms, which aim to improve the mixing of the Markov chain. In this section, we will focus on ways to extend the applicability of SGMCMC algorithms to a wider class of models. Given our choice of target \eqref{eq:target}, we have made two key assumptions, i) the parameters exist in $\st \in \mathbb{R}^d$ and ii) the potential function $U(\st)$ is a summation over independent terms. The first assumption implies that SGMCMC cannot be used to estimate $\st$ on a constrained space (e.g. $\st \in [0,1]$) and the second assumption  that our data $y_1,\ldots,y_N$ are independent or have only certain-types of dependence structure, which means that SGMCMC cannot  be applied to many time series or spatial models. We will give a short overview of some of the current research in this area.

\subsubsection*{SGMCMC sampling from constrained spaces}
\label{sec:sampl-from-constr}
%So far we have considered the problem of sampling from a target distribution with density $\pi(\st)$, where we assumed that $\st \in \mathbb{R}^d$. 

Many models contain parameters which are constrained, for example, the variance parameter $\tau^2$ in a Gaussian distribution ($\tau \in \mathbb{R}^+$), or the success probability $p$ in a Bernoulli model ($p \in [0,1]$). Simulating these constrained parameters using the Langevin dynamics \eqref{eq:Euler} will produce samples which violate their constraints, for example, if $\tau_t^2 = \st_t \gtrapprox 0$, then with high probability, $\tau_{t+1}^2<0$. One solution would be to let $h \rightarrow 0$ when $\tau^2 \rightarrow 0$, however, this would lead to poor mixing of the Markov chain near the boundary of the constrained space. A natural solution to this problem is to transform the Langevin dynamics in such a way that sampling can take place on the unconstrained space, but the choice of transformation can greatly impact the mixing of the process near the boundary. Alternatively we can project the Langevin dynamics into a constrained space \citep{brosse2017sampling,bubeck2018sampling}, however, these approaches lead to poorer non-asymptotic convergence rates than in the unconstrained setting. Recently, a mirrored Langevin algorithm \citep{hsieh2018mirrored} has been proposed, which builds on the mirrored descent algorithm \citep{beck2003mirror}, to transform the problem of constrained sampling to an unconstrained space via a mirror mapping. Unlike previous works, the mirrored Langevin algorithm has convergence rates comparable with unconstrained SGLD \citep{Dalalyan:2017}.

%Many gradient-based algorithms can be transformed to sample from general constrained spaces (examples above). However, t

The structure of some models naturally leads to bespoke sampling strategies. A popular model in the machine learning literature is the latent Dirichlet allocation (LDA) model \citep{blei2003latent}, where the model parameters are constrained to the probability simplex, meaning $\st^{(j)} \geq 0, \ j=1,\ldots,d$ and $\sum_{j=1}^d \st^{(j)} = 1$. \citet{patterson2013stochastic} proposed the first SGLD algorithm for sampling from the probability simplex. Their algorithm, stochastic gradient Riemannian Langevin dynamics (see Table \ref{tab:recipes}) allows for several transformation schemes which transform $\st$ to $\mathbb{R}^d$. However, this approach can result in asymptotic biases which dominate in the boundary regions of the constrained space. An alternative approach is to use the fact that the posterior for the LDA can be written as a transformation of independent gamma random variables. Using an alternative stochastic process instead of the Langevin diffusion, in this case the Cox-Ingersoll-Ross (CIR) process, we take advantage of the fact that its invariant distribution is a gamma distribution and apply this in the large data setting by using data subsampling on the CIR process rather than on the Langevin diffusion \citep{baker2018large}.

\subsubsection*{SGMCMC sampling with dependent data}
\label{sec:sampling-with-non}

%\PF{Things are a bit more complicated than just independent/dependent; as SGMCMC can easily be applied to simple AR models.}

Key to developing stochastic gradient MCMC algorithms is the ability to generate unbiased estimates of $\nabla U(\st)$ using data subsampling, as in \eqref{eq:U-hat-simple}. Under the assumption that data $y_i, \ i=1,\ldots,N$ are independent, the potential function $U(\st)=\sum_{i=1}^N U_i(\st)$, and its derivative, are a sum of independent terms (see Section \ref{sec:langevin-diffusion}) and therefore, a random subsample of these terms leads to an unbiased estimate of the potential function, and its derivative.
For some dependence structures, we can still write the potential as a sum of terms each of which has an $O(1)$ cost to evaluate. However for many models used for network data, time series and spatial data, using the same random subsampling approach will result in biased estimates for $U(\st)$ and $\nabla U(\st)$. To the best of our knowledge, the challenge of subsampling spatial data, such that both short and long term dependency is captured, has not been addressed in the stochastic gradient MCMC setting. For network data, an SGMCMC algorithm has been developed \citep{li2016scalable} for the mixed-member stochastic block model, which uses both the block structure of the model, and stratified subsampling techniques, to give unbiased gradient estimates. 

In the time series setting, hidden Markov models are challenging for stochastic gradient MCMC as the temporal dependence in the latent process precludes simple random data subsampling. However, such dependencies are often short range and so data points $y_i$ and $y_j$ will be approximately independent if they are sufficiently distant (i.e. $j>>i$). These properties were used by \citet{ma2017stochastic}, who proposed using SGMCMC with gradients estimated using non-overlapping, subsequences of length $2s+1$,  $\by_{i,s} = \{y_{i-s},\ldots,y_i,\ldots,y_{i+s}\}$.  In order to ensure that the subsequences are independent, \citet{ma2017stochastic} extend the length of each subsequence by adding a buffer of size $B$, to either side, i.e. $\{\by_{LB},\by_{i,s},\by_{RB}\}$, where $\by_{LB} = \{y_{i-s-B},\ldots,y_{i-s-1}\}$ and $\by_{RB}=\{y_{i+s+1},\ldots,y_{i+s+B}\}$. Non-overlapping buffered subsequences are sampled, but only $\by_{i,s}$ data are used to estimate $\hat \nabla U(\st)$. These methods introduce a bias, but one that can be controlled, with the bias often decreasing exponentially with the buffer size. This approach has also been applied to linear \citep{aicher2018stochastic} and nonlinear \citep{aicher2019stochastic} state-space models, where in the case of log-concave models, the bias decays geometrically with buffer size.

\section{Simulation Study}
\label{sec:simulation-study}

We compare empirically the accuracy and efficiency of the stochastic gradient MCMC algorithms described in Section \ref{sec:altern-stoch-grad}. We consider three popular models. Firstly, a logistic regression model for binary data classification tested on simulated data. Secondly, a Bayesian neural network \citep{neal2012bayesian} applied to image classification on a popular data set from the machine learning literature. Finally, we consider the Bayesian probabilistic matrix factorisation model \citep{salakhutdinov2008bayesian} for predicting movie recommendations based on the MovieLens data set. We compare the various SGMCMC algorithms against the STAN software \citep{Carpenter:2017}, which implements the NUTS algorithm \citep{hoffman2014no} as a method for automatically tuning the Hamiltonian Monte Carlo sampler. We treat the STAN output as the ground truth posterior distribution and assess the accuracy and computational advantages of SGMCMC against this benchmark. All of the SGMCMC algorithms are implemented using the R package \textit{sgmcmc} \citep{baker2017sgmcmc} with supporting code available online\footnote{https://github.com/chris-nemeth/sgmcmc-review-paper}.

%Secondly, an analysis of the MovieLens data set (REF) using probabilistic matrix factorisation, a factor analysis model used to identify low-rank latent features from user preference data.
\subsection{Logistic regression model}
\label{sec:logist-regr-model}

Consider a binary regression model where $\by=\{y_i\}_{i=1}^N$ is a vector of  $N$ binary responses and $\bX$ is a $N \times d$ matrix of covariates. If $\st$ is a $d-$dimensional vector of model parameters, then the likelihood function for the logistic regression model is,
$$
p(\by, \bX \ | \ \st) = \prod_{i=1}^N \left[\frac{1}{1+\exp(-\st^\top\bx_i)}\right]^{y_i}\left[1-\frac{1}{1+\exp(-\st^\top\bx_i)}\right]^{1-y_i}
$$
where $\bx_i$ is a $d-$dimensional vector for the $i$th observation. The prior distribution for $\st$ is a zero-mean Gaussian with covariance matrix $\boldsymbol{\Sigma}_{\st}=10\mathbf{I}_d$, where $\mathbf{I}_d$ is a $d \times d$ identity matrix. We can verify that the model satisfies the strong-convexity assumptions from Section \ref{sec:theory}, where $m = \lambda_{\mbox{max}}^{-1}(\boldsymbol{\Sigma}_{\st})$ and $M=\frac{1}{4}\sum_{i=1}^N \bx_i^\top\bx_i + \lambda_{\mbox{min}}^{-1}(\boldsymbol{\Sigma}_{\st})$, and $\lambda_{\mbox{min}}(\boldsymbol{\Sigma}_{\st})$ and $\lambda_{\mbox{max}}(\boldsymbol{\Sigma}_{\st})$ are the minimum and maximum eigenvalues of $\boldsymbol{\Sigma}_{\st}$.

% \PF{Is the following paragraph needed|}
% Combining the likelihood and prior distribution we can derive the gradient of the log-posterior,
% $$
% \nabla U(\st) = \sum_{i=1}^N \left[y_i\bx_i - \frac{\bx_i}{1+\exp(-\st^\top\mathbf{x_i})}\right] - \boldsymbol{\Sigma}_{\st}^{-1}\st,
% $$

We compare the various SGMCMC algorithms where we vary the dimension of $\st$, $d=\{10,50,100\}$. We simulate $N=10^5$ data points and fix the subsample size $n=0.01N$ for all test cases. % using only $1\%$ of the data at each iteration of SGMCMC. 
We simulated data under the model described above, with $\bx_i \sim N(0,\boldsymbol{\Sigma}_{\bx})$ and simulated a matrix with $\boldsymbol{\Sigma}^{(i,j)}_{\bx}=\mbox{Unif}[-\rho,\rho]^{|i-j|}$ and $\rho=0.4$. We tune the step size $h$ for each algorithm using the kernel Stein discrepancy metric outlined in Section \ref{sec:diagnostic-tests} and set the number of leapfrog steps in SG-HMC to five. We initialise each sampler by randomly sampling the first iteration $\st_0 \sim N(0,1)$.

\begin{figure}[h]
  \centering
  \includegraphics[scale=0.6]{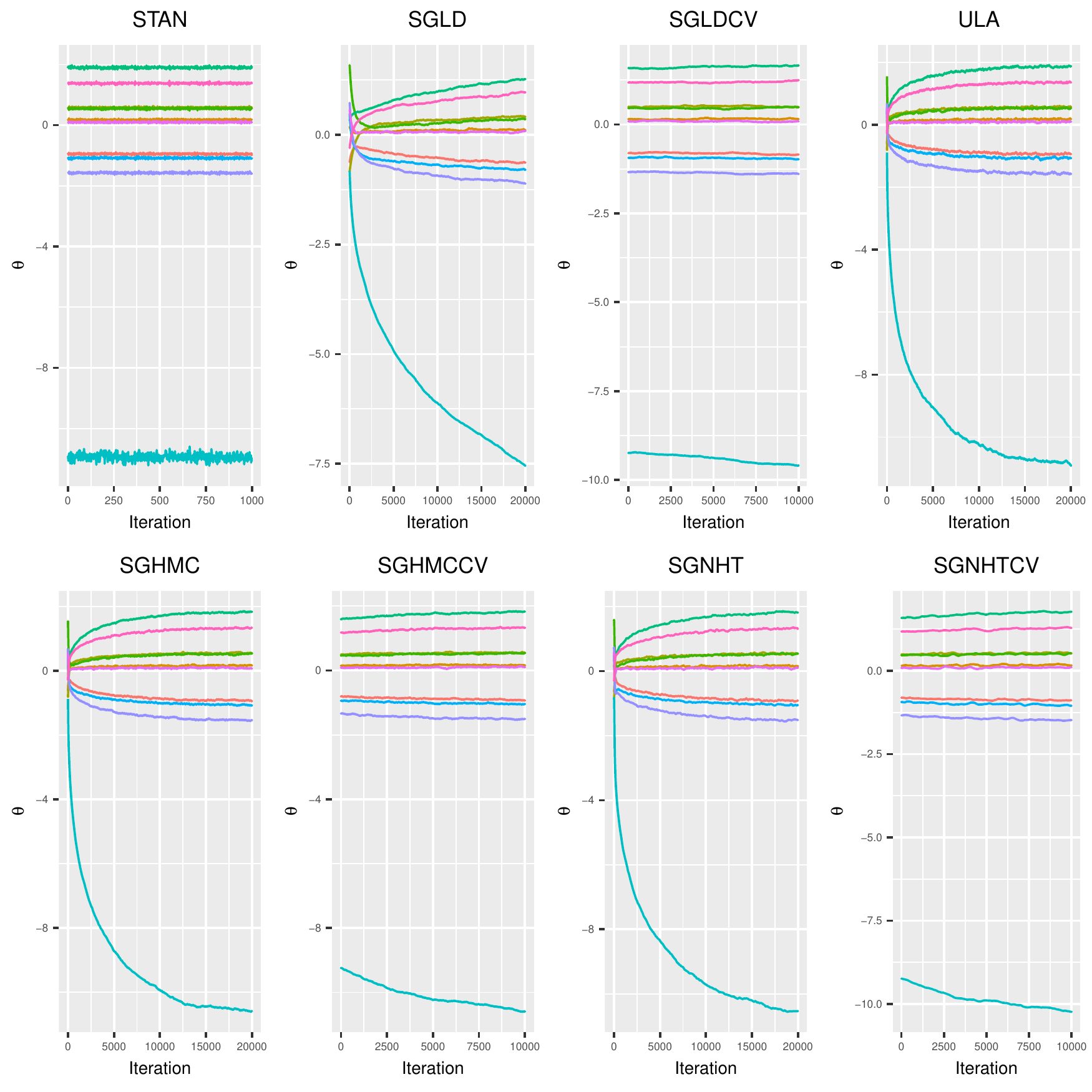}
  \caption{Trace plots for the STAN output and each SGMCMC algorithm with $d=10$ and $N=10^5$.}
  \label{fig:lregDim}
\end{figure}

For our simulations, we ran STAN for $2,000$ iterations and discarded the first $1,000$ iterations as burn-in, as these iterations are part of the algorithms tuning phase. For the SGMCMC algorithms, we ran each algorithm for $20,000$ iterations except in the case of the control variate implementations, where we ran the SGMCMC algorithm for $10,000$ iterations after iterating a stochastic gradient descent algorithm for $10,000$ iterations to find the posterior mode $\hat{\st}$. Combining the optimisation and sampling steps of the control variate method results in an equal number of iterations for all SGMCMC algorithms. Figure \ref{fig:lregDim} gives the trace plots for MCMC output of each algorithm for the case where $d=10$ and $N=10^5$. Each of the SGMCMC algorithms is initialised with the same $\st_0$ and we see that some components of $\st$, where the posterior is not concentrated around $\st_0$, take several thousand iterations to converge. Most notably SGLD, ULA, SG-HMC and SG-NHT. Of these algorithms, SG-HMC and SG-NHT converge faster than SGLD, which reflects the theoretical results discussed in Section \ref{sec:theory-sg-hmc}, but these algorithms also have a higher computational cost due to the leap frog steps (see Table \ref{tab:lregDim_metrics} for computational timings). The ULA algorithm, which uses exact gradients, also converges faster than SGLD in terms of the number of iterations, but is less efficient in terms of overall computational time. The control variate SGMCMC algorithms, SGLD-CV, SG-HMC-CV and SG-NHT-CV are all more efficient than their non-control variate counterparts in terms of the number of iterations required for convergence. The control variate algorithms have the advantage that their sampling phase is initialised at a $\st_0$ that is close to the posterior mode. In essence, the optimisation phase required to find the control variate point $\hat{\st}$ replaces the burn-in phase of the Markov chain for the SGMCMC algorithm. 

As well as the visual comparisons (Figure \ref{fig:lregDim}), we can compare the algorithms using diagnostic metrics. We use the kernel Stein discrepancy  as one of the metrics to assess the quality of the posterior approximation for each of the algorithms. Additionally, the log-loss is also a popular metric for measuring the predictive accuracy of a classifier on a held-out test data set $T_*$. In the case of predicted binary responses, the log-loss is
$$
l(\st,T_*) = -\frac{1}{|T_*|} \sum_{(y_*,\bx_*) \in T_*} y_* \log p(\bx_*,\st) + (1-y_*) \log(1-p(\bx_*,\st)),
$$
where $p(\bx_*,\st) = (1+\exp(-\st^\top\mathbf{x_*}))^{-1}$ is the probability that $y_*=1$ given covariate $\bx_*$.

Table \ref{tab:lregDim_metrics} gives the diagnostic metrics for each algorithm, where the log-loss and kernel Stein discrepancy metrics are calculated on the final $1,000$ posterior samples from each algorithm. The most notable difference between the algorithms is the computational time. Compared to STAN, all SGMCMC algorithms are between 10 to 100 times faster when $d=100$. As expected, given that STAN produces exact posterior samples, it has the lowest log-loss and kernel Stein discrepancy results. However, these results are only slightly better than the SGMCMC results and the computational cost of STAN is significantly higher. All of the SGMCMC results are similar, showing that this class of algorithms can perform well, with significant computational savings, if they are well-tuned. One of the advantages of STAN, is that the NUTS algorithm \citep{hoffman2014no} allows the HMC sampler to be automatically tuned, whereas the SGMCMC algorithms have to be tuned using a pilot run over a grid of step size values. As the step size $h$ is a scalar value, the SGMCMC samplers give an equal step size to each dimension. As discussed in Section \ref{sec:theory-example}, a scalar step size parameter will mean that the SGMCMC algorithms are constrained by the $\st$ component with the smallest variance. In Table \ref{tab:lregDim_metrics} we report the minimum effective sample for each component of $\st$ and scale this by computational time. We see that the posterior samples generated under SGMCMC algorithms have a significantly lower effective sample size compared to the STAN output, this would be improved if either the gradients were pre-conditioned \citep{ahn2012bayesian}, or the geometry of the posterior space were accounted for in the sampler (e.g. SG-RHMC), which would result in different step sizes for each component of $\st$, thus improving the overall efficiency of the sampler.

\begin{table}[h]
  \centering
  \caption{Diagnostic metrics for each SGMCMC algorithm, plus STAN, with varying dimension of $\st$ where $N=10^5$}
  \label{tab:lregDim_metrics}
    \tabcolsep=0.11cm
  \begin{tabular}{|c|c|cccccccc|}
    \hline
                & $d$ & STAN & SGLD & SGLDCV & SGHMC & SGHMCCV & SGNHT & SGNHTCV & ULA \\
    \hline
                & 10  & 21.64  & 1.74 & 1.46 & 11.24  & 6.53  & 2.56  & 1.54  &  8.05    \\
    Time (mins) & 50  & 157.24 & 2.55 & 2.06 & 13.43  & 7.76  & 3.33  & 1.93  &  29.21   \\
                & 100 & 229.76 & 3.42 & 2.60 & 16.01  & 9.63  & 4.38  & 2.36  &  51.25   \\
    \hline
    Minimum     & 10  & 119.98 & 0.43 & 0.53 & 0.42  & 0.50  & 0.32  & 0.24  & 0.29 \\
        ESS     & 50  & 28.90  & 0.25 & 0.32 & 0.19  & 0.27  & 0.11  & 0.08  & 0.19 \\
    per minute  & 100 & 10.78  & 0.22 & 0.28 & 0.16  & 0.21  & 0.10  & 0.08  & 0.17 \\

    \hline
                & 10  & 0.10 & 0.11 & 0.10 &  0.10 & 0.10 & 0.10 & 0.10 & 0.10 \\
    Log-loss    & 50  & 0.04 & 0.06 & 0.05 &  0.06 & 0.05 & 0.05 & 0.05 & 0.05 \\
                & 100 & 0.04 & 0.06 & 0.05 &  0.06 & 0.05 & 0.05 & 0.05 & 0.06 \\
    \hline
                & 10  & 6.12  & 6.26  &  6.24  & 6.18 & 6.25  & 6.21 & 6.23  & 6.19 \\
    KSD         & 50  & 9.24  & 11.73 &  11.05 & 11.59 & 11.11 & 11.00 & 11.33 & 11.30 \\
                & 100 & 11.62 & 15.70 &  15.53 & 15.64 & 15.07 & 15.14 & 15.07 & 15.97 \\  
    \hline
  \end{tabular}
\end{table}

% \begin{table}[h]
%   \centering
%   \caption{Diagnostic metrics for each SGMCMC algorithm compared against STAN with varying data size $N$, where $d=100$}
%   \label{tab:lregDim_metrics}
%   \begin{tabular}{|c|c|cccccccc|}
%     \hline
%                 & $N$ & STAN & SGLD & SGLD-CV & SG-HMC & SG-HMC-CV & SG-NHT & SG-NHT-CV & ULA \\
%     \hline
%                 & $10^3$ &   &  &  &   &   &   &   &   \\
%     Time (mins) & $10^4$ &  &  &  &   &   &   &   &    \\
%                 & $10^5$ &  &  &  &   &   &   &   &    \\
%     \hline
%                 & $10^3$ & 0.32 & 0.41 & 0.29 & 0.22 & 0.22 & 0.36 & 0.24 & 0.33 \\
%     Log-loss    & $10^4$ & 0.03 & 0.07 & 0.08 & 0.07 & 0.06 & 0.08 & 0.07 & 0.08 \\
%                 & $10^5$ & 0.04 & 0.05 & 0.05 & 0.06 & 0.05 & 0.05 & 0.05 & 0.05 \\
%     \hline
%                 & $10^3$  &  &  &  &  &  &  &  &  \\
%     KSD         & $10^4$  &  &  &  &  &  &  &  &  \\
%                 & $10^5$  &  &  &  &  &  &  &  &  \\  
%     \hline
%   \end{tabular}
% \end{table}

\subsection{Bayesian neural network}
\label{sec:bayes-neur-netw}

\begin{figure}[h]
  \centering
  \includegraphics[scale=0.4]{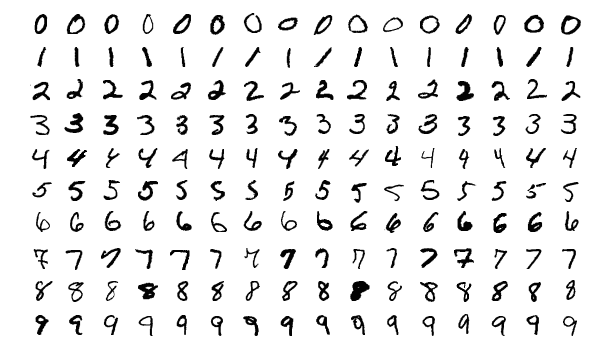}
  \caption{Sample of images from the MNIST data set taken from \url{https://en.wikipedia.org/wiki/MNIST\_database}}
  \label{fig:mnist-data}
\end{figure}

We consider the problem of multi-class classification on the popular MNIST data set \citep{Lecun2010}. The MNIST data set consists of a collection of images of handwritten digits from zero to nine, where each image is represented as $28 \times 28$ pixels (a sample of images is shown in Figure \ref{fig:mnist-data}). We model the data using a two layer Bayesian neural network with 100 hidden variables (using the same setup as \cite{chen2014stochastic}). We fit the neural network to a training data set containing $55,000$ images and the goal is to classify new images as belonging to one of the ten categories. The test set contains $10,000$ handwritten images, with corresponding labels. 

Let $y_i$ be the image label taking values $y_i \in \{0,1,2,3,4,5,6,7,8,9\}$ and $\mathbf x_i$ is the vector of pixels which has been flattened from a $28 \times 28$ image to a one-dimensional vector of length 784. If there are $N$ training images, then $\mathbf X$ is a $N \times 784$ matrix representing the full data set of pixels. We model the data as categorical variables with the probability mass function, 
\begin{align}
    p(y_i= k \, | \, \st, \mathbf x_i) = \beta_k, 
    \label{eq:beta}
\end{align}
where $\beta_k$ is the $k$th element of $\beta(\st, \mathbf x_i) = \sigma \left( \sigma \left( \mathbf x_i^\top B + b \right) A + a \right)$ and $\sigma(\bx_i)=\exp{(\bx_i)}/(\sum_{j=1}^N\exp{(\bx_i)})$ is the softmax function, a generalisation of the logistic link function. The parameters $\st = (A, B, a, b)$ will be estimated using SGMCMC, where $A$, $B$, $a$ and $b$ are matrices of dimension: $100 \times 10$, $784 \times 100$, $1 \times 10$ and $1 \times 100$, respectively. We set normal priors for each element of these parameters
\begin{align*}
    A_{kl} | \lambda_A \sim N(0, \lambda_A^{-1}), \quad B_{jk} | \lambda_B \sim N(0, \lambda_B^{-1}), \\
    a_l | \lambda_a \sim N(0, \lambda_a^{-1}), \quad b_k | \lambda_b \sim N(0, \lambda_b^{-1}), \\
    j = 1,\dots,784; \quad k = 1,\dots,100; \quad l = 1,\dots,10;
\end{align*}
where $\lambda_A, \lambda_B, \lambda_a, \lambda_b \sim \text{Gamma}(1, 1)$ are hyperparameters. 

Similar to the logistic regression example (see Section \ref{sec:logist-regr-model}), we use the log-loss as a test function. We need to update the definition of the log-loss function from a binary classification problem to the multi-class setting. Given a test set $T_*$ of pairs $(y_*, \bx_*)$, where now $y_*$ can take values $\{0-9\}$. The log-loss function in the multi-class setting is now 
\begin{equation}
  \label{eq:logLoss}
    l(\st, T_*) = - \frac{1}{|T_*|} \sum_{(y_*,\bx_*) \in T_*} \sum_{k=0}^{9} \mathbf 1_{y_{*} = k} \log \beta_k(\st, \mathbf x_*),  
\end{equation}
where $\mathbf 1_A$ is the indicator function, and $\beta_k(\st, \mb x_*)$ is the $k^{th}$ element of $\beta(\st, \mb x_*)$.

\begin{figure}[h]
  \centering
  \includegraphics[scale=0.7]{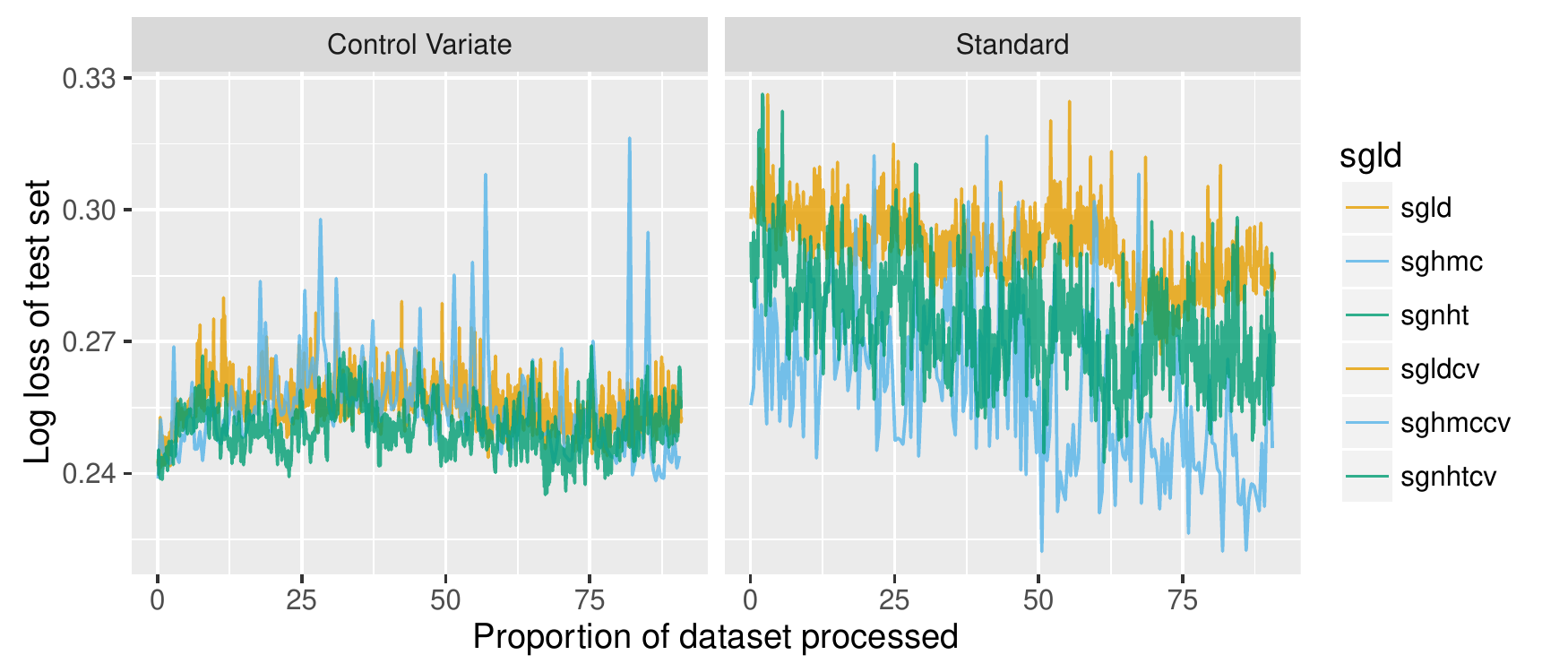}
  \caption{Log-loss calculated on a held-out test data set for each SGMCMC algorithm and its control variate version.}
  \label{fig:bnn}
\end{figure}

As in Section \ref{sec:logist-regr-model}, we compare the efficacy of the SGLD, SG-HMC and SG-NHT algorithms, as well as their control variate counterparts. We ran each of the SGMCMC algorithms for $10^4$ iterations and calculated the log-loss \eqref{eq:beta} for each algorithm. The standard algorithms have $10^4$ iterations of burn-in while the control variate algorithms have no burn-in, but $10^4$ iterations in the initial optimisation step. Note that due to the trajectory parameter $L=5$ of SG-HMC and SG-HMC-CV, these algorithms will have approximately five times greater computational cost. In order to balance the computational cost, we ran these algorithms for 2,000 iterations in order to produce comparisons with approximately equal computational time. The results are plotted in Figure \ref{fig:bnn}. As with the logistic regression examples, we note that there is some indication of improved predictive performance of the control variate methods. Among the standard methods, SG-HMC and SG-NHT have the best predictive performance, which is to be expected given the apparent trade-off between accuracy and exploration.

\subsection{Bayesian probabilistic matrix factorisation}
\label{sec:bayes-prob-matr}

% why use these models, plus movielens
Collaborative filtering is a technique used in recommendation systems to make predictions about a user's interests based on their tastes and preferences. We can represent these preferences with a matrix where the $(i,j)$th entry is the score that user $i$ gives to item $j$. This matrix is naturally sparse as not all users provide scores for all items. We can model these data using Bayesian probabilistic matrix factorisation (BPMF) \citep{salakhutdinov2008bayesian}, where the preference matrix of user-item ratings is factorised into lower-dimensional matrices representing the users' and items' latent features. A popular application of BPMF is movie recommendations, where the preference matrix contains the ratings for each movie given by each user. This model has been successfully applied to the Netflix data set to extract the latent user-item features from the historical data in order to make movie recommendations for a held-out test set of users. In this example, we will consider the MovieLens data set \footnote{https://grouplens.org/data sets/movielens/100k/} which contains $100,000$ ratings (taking values $\{1,2,3,4,5\}$) of $1,682$ movies by $943$ users, where each user has provided at least $20$ ratings. The data are already split into $5$ training and test sets ($80\%/20\%$ split) for a $5-$fold cross-validation experiment. %where the latent user-item features are inferred from the training data, and predicted movie ratings are given for each user in the test data set. 

Let $\mathbf{R} \in \mathbb{R}^{N \times M}$ be a matrix of observed ratings for $N$ users and $M$ movies where $R_{ij}$ is the rating user $i$ gave to movie $j$. We introduce matrices $\mathbf{U}$ and $\mathbf{V}$ for users and movies respectively, where $\mathbf{U}_i \in \mathbb{R}^d$ and $\mathbf{V}_j \in \mathbb{R}^d$ are $d-$dimensional latent feature vectors for user $i$ and movie $j$. The likelihood for the rating matrix is
$$
p(\mathbf{R}|\mathbf{U},\mathbf{V},\alpha) = \prod_{i=1}^N \prod_{j=1}^M \left[N(R_{ij}|\mathbf{U}_i^\top \mathbf{V}_j,\alpha^{-1})\right]^{I_{ij}}
$$

where $I_{ij}$ is an indicator variable which equals 1 if user $i$ gave a rating for movie $j$. The prior distributions for the users and movies are

$$
p(\mathbf{U}|\mathbf{\mu}_{\mathbf{U}},\Lambda_{\mathbf{U}}) = \prod_{i=1}^N N(\mathbf{U}_i|\mathbf{\mu}_{\mathbf{U}},\Lambda_{\mathbf{U}}^{-1}) \ \ \ \ \mbox{and} \ \ \ \ p(\mathbf{V}|\mathbf{\mu}_{\mathbf{V}},\Lambda_{\mathbf{V}}) = \prod_{j=1}^M N(\mathbf{V}_j|\mathbf{\mu}_{\mathbf{V}},\Lambda_{\mathbf{V}}^{-1}),
$$

with prior distributions on the hyperparameters (where $\mathbf{W}= \mathbf{U}$ or $\mathbf{V}$) given by,

$$
\mathbf{\mu}_{\mathbf{W}} \sim N(\mathbf{\mu}_{\mathbf{W}}|\mathbf{\mu}_0,\Lambda_{\mathbf{W}}) \ \ \ \ \mbox{and} \ \ \ \ \Lambda_{\mathbf{W}} \sim \mbox{Gamma}(a_0,b_0).
$$

The parameters of interest in our model are then $\st=(\mathbf{U},\mathbf{\mu}_{\mathbf{U}},\Lambda_{\mathbf{U}},\mathbf{V},\mathbf{\mu}_{\mathbf{V}},\Lambda_{\mathbf{V}})$ and the hyperparameters for the experiments are $\boldsymbol{\tau}=(\alpha,\mu_0,a_0,b_0)=(3,0,1,5)$. We are free to choose the size of the latent dimension and for these experiments we set $d=20$.

The predictive distribution for an unknown rating $R^*_{ij}$ given to movie $j$ by user $i$, is found by marginalising over the latent feature parameters
\[
p(R^*_{ij}|\mathbf{R},\boldsymbol{\tau})   = \int p(R^*_{ij}|\mathbf{U}_i,\mathbf{V}_j,\alpha) \pi(\st|\mathbf{R},\boldsymbol{\tau})  \mbox{d}\st .
\]
We can approximate the predictive density using Monte Carlo integration, where the posterior samples, conditional on the training data, are generated using the SGMCMC algorithms. The held-out test data can be used to assess the predictive accuracy of each of the SGMCMC algorithms, where we use the root mean square error (RMSE) between the predicted and actual rating as an accuracy metric.

\begin{figure}[h]
  \centering
  \includegraphics[scale=0.4]{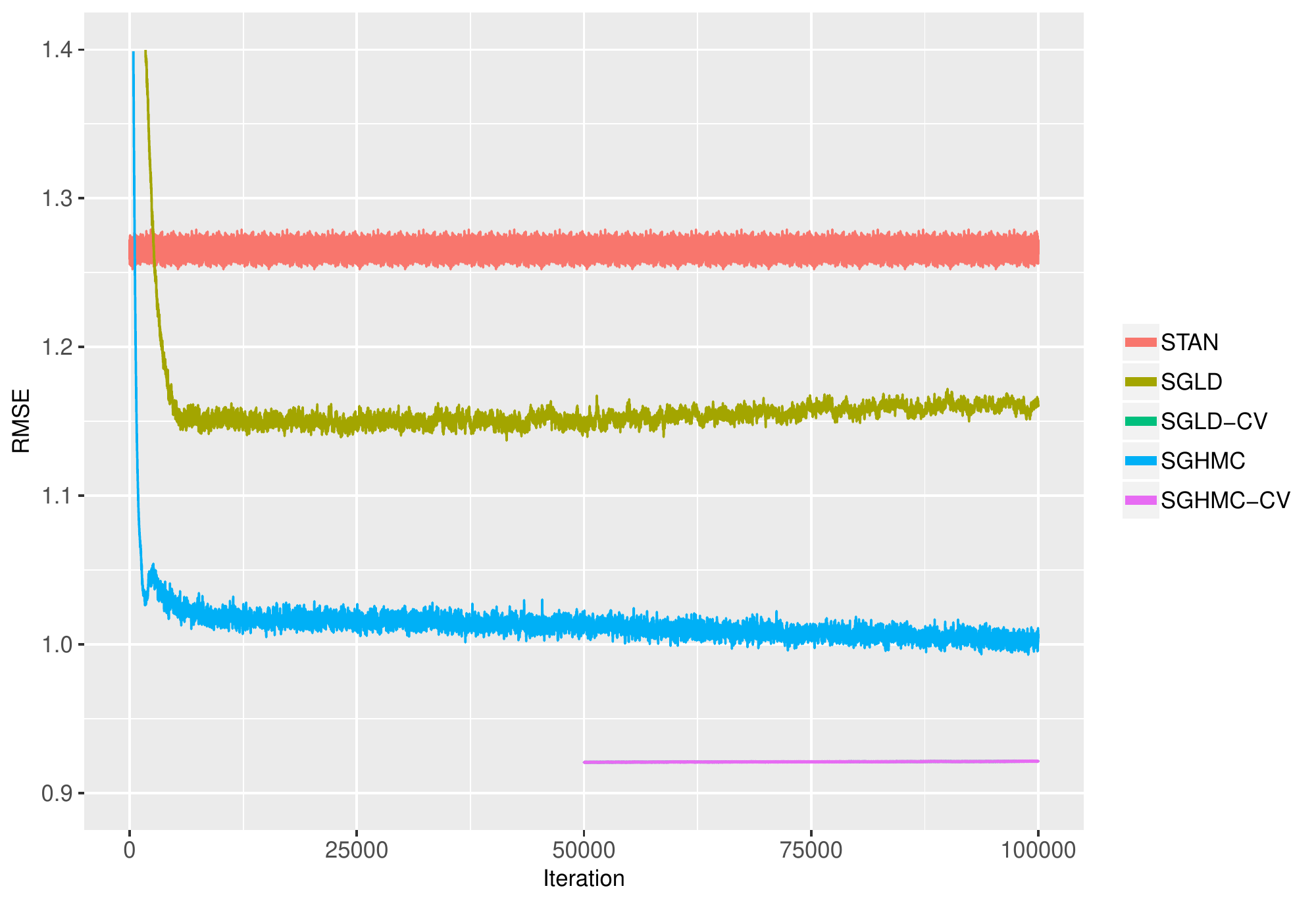}
  \caption{Root mean square error on the predictive performance of each SGMCMC algorithm averaged over five cross-validation experiments.}
  \label{fig:rmse_bpmf}
\end{figure}

We ran each of the SGMCMC algorithms for $10^5$ iterations, where for SGLD-CV and SG-HMC-CV we applied a stochastic gradient descent algorithm for $50,000$ iterations to find the posterior mode and used this as the fixed point for the control variate, as well as initialising these SGMCMC samplers from the control variate point (i.e. $\st_0=\hat{\st}$). Given the size of the parameter space, we increase the subsample size to $n=0.1N$ per iteration and tune the step size parameter for each SGMCMC algorithm using diagnostic tests (see Section \ref{sec:diagnostic-tests}) on a pilot run with $10^4$ iterations. As a baseline to assess the accuracy of the SGMCMC algorithms we applied the HMC sampler from the STAN software to the full data set and ran this for $10^4$ iterations, discarding the first half as burn-in. Figure \ref{fig:rmse_bpmf} gives the RMSE for STAN, SGLD and SG-HMC along with their control variate versions. The results show that SG-HMC produces a lower RMSE than SGLD on the test data with equally improved results for their control variate implementations. SGLD and SG-HMC quickly converge to a stable RMSE after a few thousand iterations with SGLD-CV and SG-HMC-CV producing an overall lower RMSE immediately as they are both initialised from the posterior mode, which removes the burn-in phase. Most notable from these results is that all of the SGMCMC algorithms outperform the STAN baseline RMSE. The poorer performance of STAN is attributable to running the algorithm for fewer iterations than the SGMCMC algorithms which could mean that the MCMC sampler has not converged. Running STAN for 10\% of the iterations of the SGMCMC algorithms took 3.5 days, whereas SGLD, SGLD-CV, SG-HMC and SG-HMC-CV took 3.1, 3.5, 16.4 and 16.8 hours, respectively. Therefore, while SGMCMC algorithms produce biased posterior approximations compared to exact algorithms, such as STAN, they can produce accurate estimates of quantities of interest at significantly reduced computational cost.

\section{Discussion}
\label{sec:discussion}

% \PF{This first paragraph reads more like an intro paragraph than discussion.}
% In recent years, significant technological developments, most notably the Internet, have provided the means to record and store tremendous quantities of data from almost every aspect of our daily lives. This has provided statisticians and data scientists with the opportunity to build evermore sophisticated models to better understand the processes that generate data and make predictions with greater accuracy. A key challenge when fitting a statistical model is inferring the model's parameters. If we are to take a Bayesian approach to statistical modelling, then developing increasingly complex models naturally leads to non-conjugate posterior distributions which do not admit a closed-form solution. As a result of its favourable theoretical properties, Markov chain Monte Carlo algorithms have become the workhorse of Bayesian inference and are now widely accepted in most scientific fields. At best, the computational cost of using MCMC scales linearly with the size of the data set, but as large data sets become increasingly common, traditional MCMC algorithms are failing to keep pace with the data revolution.

In this paper we have provided a review of the growing literature on stochastic gradient MCMC algorithms. These algorithms utilise data subsampling to significantly reduce the computational cost of MCMC. As shown in this paper, these algorithms are theoretically well-understood and provide parameter inference at levels of accuracy that are comparable to traditional MCMC algorithms. Stochastic gradient MCMC is still a relatively new class of Monte Carlo algorithms compared to traditional MCMC methods and there remain many open problems and opportunities for further research in this area.

Some key areas for future development in SGMCMC include:
\begin{itemize}
\item  New algorithms - as discussed in Section \ref{sec:theory-sg-hmc}, SGMCMC represents a general class of scalable MCMC algorithms with many popular algorithms given as special cases, therefore it is possible to derive new algorithms from this general setting which may be more applicable for certain types of target distribution.
\item  General theoretical results - most of the current theoretical results which bound the error of SGMCMC algorithms assume that the target distribution is log-concave. Relaxing this assumption could lead to similar non-asymptotic error bounds for a broader class of models, for example, in the case of multimodal posterior distributions.  
\item Tuning techniques - as outlined in Section \ref{sec:diagnostic-tests}, the efficacy of SGMCMC is dependent on how well the step size parameter is tuned. Standard MCMC tuning rules, such as those based on acceptance rates, are not applicable and new techniques, such as the Stein discrepancy metrics, can be computationally expensive to apply. Developing robust tuning rules, which can be applied in an automated fashion, would make it easier for non-experts to use SGMCMC methods in the same way that adaptive HMC has been applied in the STAN software.
\end{itemize}

A major success of traditional MCMC algorithms, and their broad appeal in a range of application areas, is partly a result of freely available software, such as WinBUGS \citep{lunn2000winbugs}, JAGS \citep{plummer2003jags}, NIMBLE \citep{de2017programming} and STAN \citep{Carpenter:2017}. Open-source MCMC software, which may utilise specials features of the target distribution, or provide automatic techniques to adapt the tuning parameters, make MCMC methods more user-friendly to general practitioners. Similar levels of development for SGMCMC, which provide automatic differentiation and adaptive step size parameter tuning, would help lower the entry level for non-experts. Some recent developments in this area include \textit{sgmcmc} in R \citep{baker2017sgmcmc} and \textit{Edward} in Python \citep{tran2016edward}, but further development is required to fully utilise the general SGMCMC framework.

% big data
% better models
% step change
% challenges of inference - mcmc not designed for big data, growth in approximate methods, e.g. VB
% sgmcmc has a lot of advantages - speed and theoretical support

\section*{Acknowledgements}
\label{sec:acknowledgments}

The authors would like to thank Jack Baker for his guidance and help using the \texttt{sgmcmc} package. CN gratefully acknowledges the support of EPSRC grants EP/S00159X/1 and EP/R01860X/1. PF is supported by the EPSRC-funded projects Bayes4Health EP/R018561/1 and CoSInES EP/R034710/1.

\bibliographystyle{royal}
\bibliography{sgmcmc}

\end{document}